\documentclass[12pt]{article}
\usepackage{amsfonts}
\usepackage{amsmath}
\usepackage{amssymb}
\usepackage{mathptmx}
\usepackage{geometry}
\usepackage{graphicx}
\usepackage{hyperref}
\usepackage{cancel}
\usepackage{textcomp}
\usepackage{tikz}
\usepackage{bm}
  \usepackage{mathrsfs}
  \usepackage{filecontents}
\usepackage{times}
\usepackage{epsfig}
\usepackage{xcolor}
\usepackage{slashed}
\usepackage{booktabs}% build a table
\usepackage{latexsym}
\usepackage{verbatim}
\usepackage{extarrows}
\usepackage{multirow}
\usepackage{rotating}
\usepackage{colortbl}
\usepackage{indentfirst}
\usepackage[numbers,sort&compress]{natbib}
\usepackage{soul}
\definecolor{mygray}{gray}{.9}
\definecolor{intnull}{RGB}{213,229,255}
\usepackage{arydshln}
\usepackage{diagbox}
\usepackage{caption}
\usepackage{tikz}
\usetikzlibrary{arrows,shapes,chains}
\usepackage{graphicx, subfig}
\begin{document}
\renewcommand{\thefootnote}{\fnsymbol{footnote}}
\baselineskip=16pt
\pagenumbering{arabic}
\vspace{1.0cm}
\begin{center}
{\Large\sf Internal structure and its connection with thermodynamics and dynamics in black holes}
\\[10pt]
\vspace{.5 cm}

{Yan-Gang Miao\footnote{Corresponding author. E-mail address: miaoyg@nankai.edu.cn} and Hao Yang\footnote{E-mail address: yanghao4654@mail.nankai.edu.cn}}
\vspace{3mm}

{School of Physics, Nankai University, Tianjin 300071, China}

\vspace{4.0ex}
\end{center}
\begin{center}
{\bf Abstract}
\end{center}

\renewcommand{\thefootnote}{\arabic{footnote}}
%In general, a singular metric leads to an incomplete spacetime, while a non-singular metric gives rise to a complete (an incomplete) spacetime if the corresponding Ricci and Kretschmann scalars converge (diverge) and radial geodesics are complete (incomplete) at the center of a black hole.
In general, a finite metric function at the center of a black hole describes a non-singular spacetime  but an infinite metric at the center gives a singular spacetime, where the former is associated with convergent Ricci and Kretschmann scalars together with complete radial geodesics, while the latter is related to divergent Ricci and Kretschmann scalars together with incomplete radial geodesics.
For the charged black hole in the four-dimensional Einstein-Gauss-Bonnet theory, we find that its metric function is finite at its center in  one region of parameters but complex in the other region of parameters. The finite case describes a strange spacetime which presents the Ricci and Kretschmann scalars of a singular spacetime and the radial geodesics of a non-singular spacetime, while the complex case gives rise to the similar situation. We verify that the cosmic censorship conjecture is maintained for the black hole model. 
%{\color{red}which leads a complete spacetime, even though the corresponding scalars diverge at the center.}
Further, we investigate the second-order phase transition, quasinormal modes of perturbation of a test scalar field in the eikonal limit,  and shadow radius for the black hole and find that the thermodynamic and dynamic properties depend on the metrics. In this way, we connect the internal structure with the thermodynamics and dynamics for this black hole.
Moreover, we compare such a black hole of modified gravity with the Reissner-Nordstr\"om black hole of Einstein's general relativity in these thermodynamic and dynamic properties.
%The resulting effects might be observed in future experiments when compared with those of the Reissner-Nordstr\"om black hole of Einstein's general relativity.
%and compare the results with that of the classical Reissner-Nordstr\"om black hole.
%Furthermore, differences in observables such as black hole  and black hole  are taken into consideration.
%The different ranges induce the differences of characteristic variables in thermodynamics and dynamics.
\newpage

%%%%%%%%%%%%%%%%%%%%%%%%%%%%%%%%%%%%%%%%%%%%%%%%%%%%%%%%%%%%%%%%%%%%%%
\section{Introduction}
\label{sec:intr}
%%%%%%%%%%%%%%%%%%%%%%%%%%%%%%%%%%%%%%%%%%%%%%%%%%%%%%%%%%%%%%%%%%%%%%
%%%%%%%%%%%%%%%%%%%%%%%%%%%%%%%%%%%%%%%%
Traditionally, Lovelock's theorem has shown~\cite{A1} that Einstein's general relativity is the unique theory of gravity in the four-dimensional spacetime if the diffeomorphism invariance, metricity, and second order equations of motion are required. However, from the point of view of conformal gravity, the Einstein-Gauss-Bonnet (EGB) theory in four dimensions, the so-called 4D EGB gravity, was given~\cite{X1} with conformal anomaly and developed~\cite{X2} with Gauss-Bonnet entropic corrections.
%{\color{red}However, was proposed in gravity theory with conformal anomaly, and its development with the cosmological constant.}
Recently, it was rediscovered~\cite{P2} by the way of starting with a higher dimensional EGB theory, rescaling the EGB parameter and taking the limit of the spacetime dimension to four.
%Correspondingly, its static spherically symmetric solution, the uncharged 4D EGB black hole has been given, {\color{red}which is same as the solution in gravity theory with conformal anomaly.}
%It is worth noting that the similar solutions had been obtained in the research of gravity theory with conformal anomaly~\cite{X1} and with Gauss-Bonnet entropic corrections~\cite{X2}.
Later, a lot of related studies have been done, which can roughly be classified into three types. The first type of research mainly focuses on the extension of the 4D EGB gravity, such as constructing the charged black hole~\cite{P1}, the Bardeen black hole~\cite{P3}, the noncommutative inspired black hole~\cite{P4}, and the Born-Infeld black hole~\cite{P5}, {\em etc}. The second type of the relevant studies argues the self consistency of the strategy adopted in Ref.~\cite{P2} and of the theory itself on various aspects, for instance, field equations and Lagrangians~\cite{P6}, instability of vacua~\cite{FWSu}, and some other criticisms~\cite{C1}. Furthermore, a well-defined 4D gravity theory was proposed~\cite{ADD1,ADD2} from the Horndeski scalar-tensor theory and from a minimally modified gravity theory, respectively.
The last type of research reveals~\cite{D1} new interesting properties of the 4D EGB gravity in dynamics and thermodynamics of black holes, in particular, quasinormal modes and shadows~\cite{B1}, and restrictions to the EGB parameter from the observations of M87*~\cite{X3} and the speed of gravitational waves measured by GW170817 and GRB 170817A~\cite{X4}, and from validity of the weak cosmic censorship conjecture~\cite{X5}, {\em etc}.

The charged 4D EGB black hole can be divided~\cite{X1,P1} into two classes depending on the positivity or negativity of the EGB parameter,\footnote[1]{We make a note that the EGB parameter is positive or negative in different cases. If the causality of the bulk structure in the AdS space is imposed~\cite{GeSin}, the EGB parameter should be negative; if no superluminal sound speed of gravitational waves is required~\cite{CLWY} in the stochastic gravitational wave background, the EGB parameter should be positive.} where the metric function with the positive parameter is complex at the center, while the metric function with the negative parameter is real and finite when the negative parameter and the charge satisfy a certain  constraint.
In general, if a metric function is finite at the center of a black hole, it describes a non-singular spacetime  in which the Ricci and Kretschmann scalars are convergent and the radial geodesics are complete; but if a metric function is infinite at the center, it gives a singular spacetime in which the Ricci and Kretschmann scalars are divergent and the radial geodesics are incomplete.
Here we point out that the charged 4D EGB black hole is an exception.
The metric function of this black hole is finite everywhere for the negative EGB parameter plus its constraint to the charge, but it describes a strange spacetime in which the Ricci and Kretschmann scalars are divergent, however, the radial geodesics are complete, which is quite amazing because the Ricci and Kretschmann scalars present the properties of a singular spacetime but the radial geodesics show the properties of a non-singular spacetime.
For the positive EGB parameter, the metric function of this black hole takes a complex value at its center, and the resulting spacetime has similar properties of curvature invariants and radial geodesics to those in the case of the negative EGB parameter.

%ask an interesting question, how a black hole presents its features in thermodynamics and dynamics in the two cases, where one case corresponds to a singular metric associated with a certain region of parameters and the other case to a non-singular metric associated with another certain region of parameters.
%and both metrics describe an incomplete spacetime.
%As such a spacetime region is always hidden by an event horizon, we cannot judge just from geometry.
%Thanks to the thermodynamics and dynamics of black holes, we try to answer this question by investigating the thermodynamic and dynamic differences caused by the metrics with or without singularity.
%further investigate constraints to the EGB parameter from the point of view of the completeness of spacetime.
%and find a contradiction between the completeness of spacetime and the smoothness of limiting from the charged to uncharged cases of 4D EGB black holes.
Due to the peculiarity of the charged 4D EGB black hole, we further investigate how the finite and complex metric functions result in different thermodynamic and dynamic effects. We focus on the second-order phase transition, the quasinormal modes of perturbation of a test scalar field in the eikonal limit,  and shadow radius. In addition, considering that the charged
4D EGB black hole turns back to the Reissner-Nordstr\"om (RN) black hole in the limit of vanishing EGB parameter, we compare the charged 4D EGB black hole with the RN black hole in these thermodynamic and dynamic properties and then obtain some interesting relations.

The second-order phase transition point of black holes was first found by Daives~\cite{P7}, known as ``the Davies point". When a black hole evolves past the Davies point, the sign of its heat capacity changes, indicating that the black hole has undergone a phase transition. As the heat capacity diverges at the Davies point, the phase transition is second order.
The Davies points exist in the RN black hole and the Kerr black hole, but not in the Schwarzschild black hole whose heat capacity is always negative. We pointed out~\cite{P13} that the Davies point appears in the uncharged 4D EGB black hole. Here we shall further investigate the behaviors of Davies points for the charged 4D EGB black hole.

%The observation of gravitational waves from a binary black hole merger was first reported by the LIGO Scientific and Virgo collaborations~\cite{P9} in 2016. Therefore, the gravitational wave opens a new window for us to study the properties of black holes.
Quasinormal modes provide~\cite{P15} a characteristic variable of dynamics to depict the ringdown phase of black hole mergers, where  their real parts correspond to the oscillating frequency of perturbations, while their imaginary parts to the damping time.
%They depend on metrics, so the singularity of metrics will be reflected by the quasinormal modes.
Normally the quasinormal modes can only be computed numerically, but in the eikonal limit they can be derived~\cite{P10} analytically, {\em i.e.}, the light ring/quasinormal mode correspondence, where the angular velocity of unstable null geodesics determines the real part of the quasinormal modes and the Lyapunov index does the imaginary part.
Although this correspondence is not valid~\cite{P14} for all field perturbations, it holds for our test scalar field.
Besides the quasinormal modes, another important observable is the shadow radius of a black hole~\cite{P12} which is inversely proportional to the real part of the quasinormal modes of perturbation of a test scalar field in the eikonal limit when observed at infinity.
Thus, the quasinormal modes in the eikonal limit and the shadow radius represent the dynamic properties of the charged 4D EGB black hole.

This paper is organized as follows. In Sec.~\ref{sec: Charged}, we determine the constraints under which the metric function of the charged 4D EGB black hole is real and finite or complex at the center, and discuss the cosmic censorship conjecture for this black hole model.
Next, we examine in Sec.~\ref{sec:the Davies point} the behaviors of Davies points when the EGB parameter takes values in different ranges. We continue our discussions in Sec.~\ref{sec:QNM} on the quasinormal modes of perturbation of a test scalar field in the eikonal limit. Finally,
%we discuss  a charged 4-dimensional EGB black hole under different ranges of the EGB parameter, and compare with the RN black hole to calculate the theoretical maximum relative deviation of observables.
we give our conclusions in Sec.~\ref{sec:conclusions} where some comments and further extensions are included. We note that the term ``complex metric function" only means the property at the center of black holes throughout this paper.

%%%%%%%%%%%%%%%%%%%%%%%%%%%%%%%%%%%%%%%%%%%%%%%%%%%%%%%%%%%%%%%%%%%%%%
\section{Charged 4D EGB black hole}
\label{sec: Charged}
%%%%%%%%%%%%%%%%%%%%%%%%%%%%%%%%%%%%%%%%%%%%%%%%%%%%%%%%%%%%%%%%%%%%%%
%%%%%%%%%%%%%%%%%%%%%%%%%%%%%%%%%%%%%%%%
The metric function of the charged 4D EGB black hole  takes~\cite{X1,P1} the form,
\begin{equation}
f_\alpha(r)=1+\frac{r^2}{2\alpha}\left[1-\sqrt{1+4\alpha\left(\frac{2M}{r^3}-\frac{q^2}{r^4}\right)}\right],\label{metric}
\end{equation}
where $\alpha$ stands for the EGB parameter, $M$ the mass, and $q$ the charge.
Under the limit $\alpha\rightarrow 0$, the charged 4D EGB black hole turns back to the RN black hole with the metric function as follows,
\begin{equation}
f_{\rm RN}(r)=1-\frac{2M}{r}+\frac{q^2}{r^2}.\label{metric_RN}
\end{equation}
By solving the algebraic equation, $f_\alpha(r)=0$, one can obtain the event horizons of the charged 4D EGB black hole,
\begin{equation}
\label{eq:horizon}
r_{\rm H}^\pm=M\pm\sqrt{M^2-q^2-\alpha}.
\end{equation}
The existence of horizons requires
\begin{equation}
	M^2-q^2-\alpha\geq0.\label{conshor}
\end{equation}

\subsection{Complex metric function}

When $r\rightarrow 0$, the asymptotic behavior of the term under the square root in Eq.~(\ref{metric}) is
\begin{equation}
1+4\alpha\left(\frac{2M}{r^3}-\frac{q^2}{r^4}\right)\longrightarrow -\frac{4\alpha q^2}{r^4},
\end{equation}
which means that the metric function takes a complex value if $\alpha >0$. %we follow the procedure in \cite{P2} and choose , .
%So instead of having a singularity at the center of the black hole we have a singular region, where the metric is complex.
We notice that the complex metric function is different from the usual function that is infinite at the center of black holes. The appearance of a complex metric function is parameter dependent. Next, we investigate the Ricci scalar, the Kretschmann scalar, and  the radial geodesic in the case of the positive EGB parameter.

\subsubsection{Ricci scalar and Kretschmann scalar}

%It is obvious that the spacetime of charged 4D EGB black holes {\color{red}has a singularity} whether the EGB parameter is positive or negative, which can be seen clearly from the asymptotic form of the Ricci scalar in the limit of $r\to 0$, $\lim_{r\rightarrow 0}R\sim \frac{32q^4\alpha}{r^8}$.

It is well known that the divergence of Ricci and Kretschmann scalars  reflects the existence of an irreducible singularity in spacetime. So it is obvious that the spacetime of charged 4D EGB black holes has singularity when the EGB parameter is positive, which can be seen clearly from the asymptotic form of the Ricci scalar in the limit of $r\to 0$,
\begin{equation}
\lim_{r\rightarrow 0}R\sim \frac{4iq}{r^2\sqrt{\alpha}},\label{riccige0}
\end{equation}
and the asymptotic form of the Kretschmann scalar in the limit of $r\to 0$,
\begin{equation}
\lim_{r\rightarrow 0}K\sim -\frac{4q^2}{r^4\alpha}.\label{kretge0}
\end{equation}

\subsubsection{Radial geodesic}
For a spherically symmetric spacetime with the metric function $f(r)$, the radial time-like geodesic of a neutral particle satisfies~\cite{AD1}
\begin{equation}
  \left(\frac{dr}{d\tau}\right)^2+f(r)=E^2,
\end{equation}
where $\tau$ is the proper time and $E$ the total energy per unit mass, and $E^2=1$ corresponds to a free particle. Thus the square of radial velocities associated with $\tau$ of a free neutral  particle is
\begin{equation}\label{velocity}
v^2\equiv\left(\frac{dr}{d\tau}\right)^2=1-f(r).
\end{equation}

First of all, let us review the incompleteness of the radial geodesic~\cite{M1} in the uncharged 4D EGB black hole spacetime.
The square of radial velocities of a neutral particle falling into the center of the black hole is
\begin{equation}
\left(\frac{dr}{d\tau}\right)^2=E^2-1+\sqrt{\frac{2M}{\alpha}}r^{1/2}+O(r^{3/2}).
\end{equation}
Thus, this square is exactly zero when a free neutral particle reaches the center, resulting in the spacetime incompleteness.

Next, we calculate the square for a neutral particle falling into the center of the charged 4D EGB black hole when $\alpha>0$,
%Since in the charged case, the velocity of the particle near the origin is
%\begin{equation}
%\left(\frac{dr}{d\tau}\right)^2=E^2-1-\frac{r^2}{2\alpha}\left[1-\sqrt{1+4\alpha\left(\frac{2M}{r^3}-\frac{q^2}{r^4}\right)}\right].
%\end{equation}
\begin{equation}
\left(\frac{dr}{d\tau}\right)^2=E^2-1-\sqrt{-\frac{q^2}{\alpha}}+\sqrt{-\frac{q^2}{\alpha}}\cdot\frac{M}{q^2}r+O(r^2).\label{sqrvel4degb}
\end{equation}
We can see that this square is imaginary for a free neutral particle ($E^2=1$) when $r\to 0$.
This shows that no physical observer can reach the black hole center in a finite proper time.
In addition, we compute the innermost radius that the particle can reach,
\begin{equation}
r_{\rm in}=\frac{q^2}{2M},\label{innerradius}
\end{equation}
which is bigger than the critical radius $r_b$ of the zone of the complex metric function.\footnote[2]{According to Eq.~(\ref{metric}), the critical radius of the zone of the complex metric function satisfies the equation: $ r_b^4+8\alpha Mr_b-4\alpha q^2=0$, and the only real and positive solution can be obtained, $ r_b= \sqrt{\frac{2 \sqrt{2} \alpha  M}{\sqrt{\sqrt[3]{{4}/{9}} A-{2 \sqrt[3]{{2}/{3}} \alpha  q^2}/{A}}}-\left(\sqrt[3]{{4}/{9}} A/2-{ \sqrt[3]{{2}/{3}} \alpha  q^2}/{A}\right)}
-\sqrt{\sqrt[3]{{4}/{9}} A/2-{ \sqrt[3]{{2}/{3}} \alpha  q^2}/{A}}$, where $A\equiv \sqrt[3]{\sqrt{3\left(27 \alpha ^4 M^4+4 \alpha ^3 q^6\right)}+9 \alpha ^2 M^2}$. It is not easy to determine which is bigger, $r_b$ or $r_{\rm in}$.  To this end, we introduce the function, $s(r)\equiv r^4+8\alpha Mr-4\alpha q^2$, and investigate its monotonicity. Because the derivative of $s(r)$ with respect to $r$ is positive when $\alpha>0$, i.e., $ {ds(r)}/{dr}=4r^3+8\alpha M>0$, $s(r)$ is a monotonically increasing function. Further, considering $s(r_{\rm in})=\left(\frac{q^2}{2M}\right)^4$ and $s(r_b)=0$, that is, $s(r_{\rm in})>s(r_b)$, thus we deduce $r_{\rm in}>r_b$.} Note that the zone within the innermost radius is unreachable,  which implies that the zone of the complex metric function is unreachable, too.

\subsection{Finite metric function}\label{sec2.2}

We continue our analyses of the metric function for the charged 4D EGB black hole by combining the function with the event horizons.
At first, in order to have a real metric, we impose the constraint,
\begin{equation}
\label{eq: main constraint}
1+4\alpha\left(\frac{2M}{r^3}-\frac{q^2}{r^4}\right)\geq 0,\qquad r\in [0,+\infty).
\end{equation}
When $r>\frac{q^2}{2M}$, meaning $\frac{2M}{r^3}-\frac{q^2}{r^4}>0$, the above condition becomes
\begin{equation}
\label{eq: constraint 1}
\alpha\geq -\frac{r^4}{4(2Mr-q^2)}\equiv h(r).
\end{equation}
Because $h(r)$ takes its maximum at $r_0=\frac{2q^2}{3M}$,
\begin{equation}
h(r_0)=-\frac{4q^6}{27M^4},
\end{equation}
Eq.~(\ref{eq: constraint 1}) gives the lower bound of $\alpha$,
\begin{equation}
\alpha\geq-\frac{4q^6}{27M^4}.\label{firsec}
\end{equation}
Moreover, when $0<r<\frac{q^2}{2M}$, meaning $\frac{2M}{r^3}-\frac{q^2}{r^4}<0$, Eq.~(\ref{eq: main constraint}) implies
\begin{equation}
\label{eq:constraint2}
\alpha\leq h(r).
\end{equation}
Considering
\begin{equation}
\frac{\mathrm{d} h(r)}{\mathrm{d} r}=\frac{r^3(2q^2-3Mr)}{2q^2-2Mr^2}>0,
\end{equation}
and $h(r)\rightarrow 0$ when $r\rightarrow 0$, we can see that $h(r)$ increases monotonically in the range of $0<r<\frac{q^2}{2M}$ and its minimum is zero. So, Eq.~(\ref{eq:constraint2}) gives the upper bound of $\alpha$,
\begin{equation}
\alpha\leq 0.\label{seccons}
\end{equation}
Combining Eq.~(\ref{firsec}) with Eq.~(\ref{seccons}), we get the constraint to $\alpha$,
\begin{equation}
-\frac{4q^6}{27M^4}\leq \alpha\leq 0, \label{consmetr}
\end{equation}
which is just the requirement to maintain a real metric function.

In addition, we need to consider the restrictions from horizons, i.e. the existence of horizons, Eq.~(\ref{conshor}).
For the case of $q^2\leq M^2$, Eq.~(\ref{conshor}) does not give extra restrictions to Eq.~(\ref{consmetr}), while for the case of $q^2>M^2$, the combination of Eq.~(\ref{conshor}) and Eq.~(\ref{consmetr}) leads to a more compact condition than Eq.~(\ref{consmetr}),
\begin{equation}
-\frac{4q^6}{27M^4}\leq \alpha\leq M^2-q^2.\label{consmetr2}
\end{equation}
Note that we shall verify this inequality soon later by checking $-\frac{4q^6}{27M^4}\leq M^2-q^2$.

For the sake of concision in the following demonstrations, we introduce the rescaling of variables,
\begin{equation}
\frac{q}{M}\rightarrow Q,\qquad \frac{\alpha}{M^2}\rightarrow a,\label{rescale}
\end{equation}
with which we rewrite the constraint conditions that consist of Eq.~(\ref{consmetr}) and Eq.~(\ref{consmetr2}) in a dimensionless form,
\begin{subequations}
\begin{equation}
-\frac{4}{27}Q^6 \leq a \leq 0, \qquad \text{for}\quad Q^2\leq1,\label{constcom}
\end{equation}
\begin{equation}
-\frac{4}{27}Q^6 \leq a \leq 1-Q^2, \qquad \text{for}\quad Q^2>1. \label{constcom2}
\end{equation}
\end{subequations}

\begin{figure}[h!]
    \centering
    \includegraphics[width=0.6\textwidth]{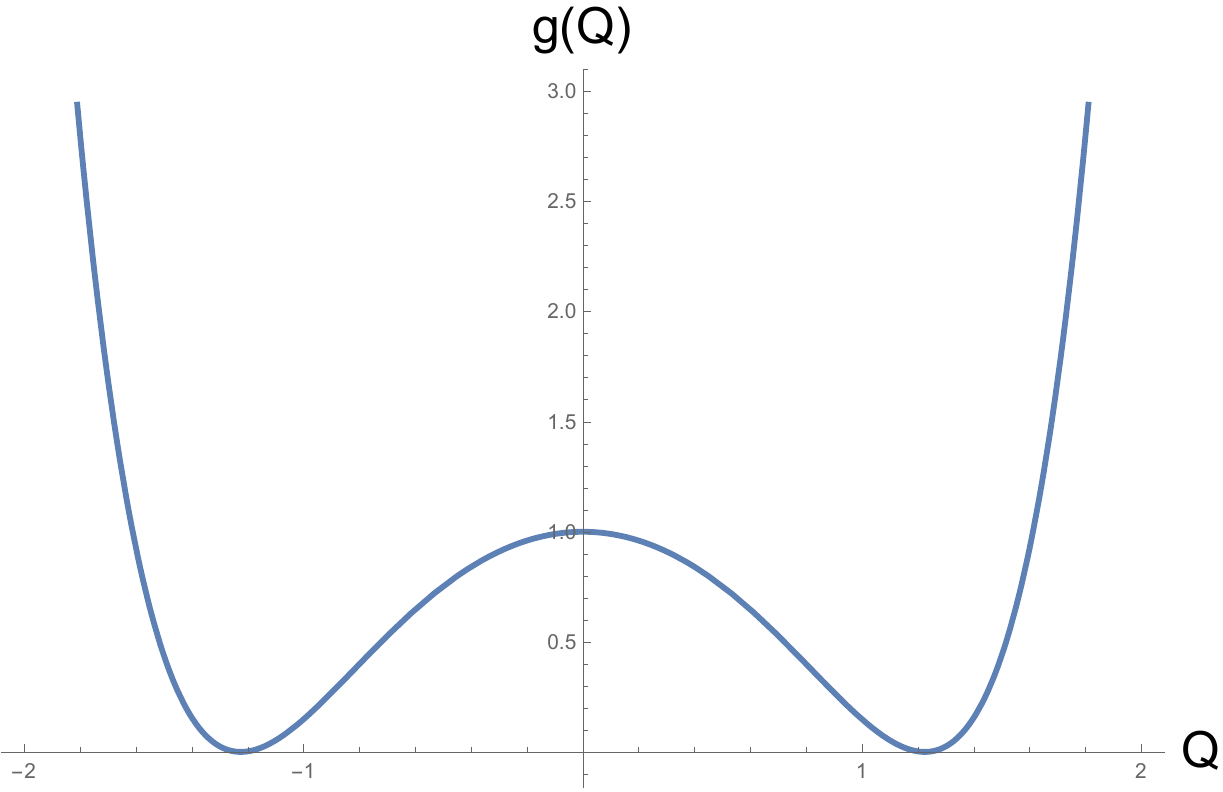}
    \caption{The function $g(Q)$ is tangent to the horizontal axis $Q$ at the two points, $(\pm\sqrt{\frac{3}{2}},0)$.}
    \label{Fig: A function of Q}
\end{figure}

Now we compensate the proof for the inequality Eq.~(\ref{consmetr2}) or its dimensionless form Eq.~(\ref{constcom2}), that is, to check $1-Q^2\geq-\frac{4}{27}Q^6$. To this end, we define a function,
\begin{equation}
g(Q)\equiv 1-Q^2+\frac{4}{27}Q^6,
\end{equation}
and then prove $g(Q)\geq 0$ for $Q^2>1$.
We draw the graph of $g(Q)$ with respect to $Q$ in Fig.~\ref{Fig: A function of Q}, from which we can finish our proof.
We pay attention to the two zero points of $g(Q)$ located at $Q=\pm\sqrt{\frac{3}{2}}$, where $a$ has one unique value, $a=-\frac{1}{2}$. This means that the charged 4D EGB black hole has only one horizon when the rescaled EGB parameter equals $-\frac{1}{2}$, i.e. the inner and outer horizons merge.

Besides the rescaled charge and EGB parameter, see Eq.~(\ref{rescale}), we rescale the radial coordinate $r$ by
\begin{equation}
\frac{r}{M}\rightarrow x,\label{rescalex}
\end{equation}
%\begin{figure}[h!]
   %\centering
    %\includegraphics[width=0.45\textwidth]{region_a.pdf}
    %\caption{The horizontal axis corresponds to the resacling charge parameter $Q$, and the vertical axis corresponds to the resacling EGB parameter $a$. In the figure, the shadow part means the region allowed by constraint conditions, and there is only outer horizon in the deeper color part.}
   % \label{Fig: The region of $a$}
%\end{figure}
and then rewrite the horizons Eq.~(\ref{eq:horizon}) as
\begin{equation}
x_{\rm H}^\pm=1\pm\sqrt{1-Q^2-a},\label{dlhor}
\end{equation}
where it seems that $a$ takes the range from minus infinity to zero.
%We have discussed the relationship between horizons and the upper limit of $\alpha$ in above analysis. Now let's talk about the relationship between the lower limit of $a$ and the event horizon of the black hole. When $\alpha=-\frac{4q^6}{27M^4}$, horizons become
%\begin{equation}
%x_H^\pm=1\pm\sqrt{1-Q^2+\frac{4}{27}Q^6}=1\pm\sqrt{g(Q)}
%\end{equation}
%\begin{figure}[h!]
   % \centering
   % \includegraphics[width=0.45\textwidth]{horizons_Q.pdf}
    %\caption{The relationship between $Q$ and horizons in the inferior limit of $\alpha$.}
    %\label{Fig: The relationship between horizons in the inferior limit of $\alpha$ and $Q$}
%\end{figure}
%When $g(Q)>1\Rightarrow Q^2>\frac{3\sqrt{3}}{2}$, $x_H^-<0$, and there is only one horizon in this situation. In summary, there are two horizons (inner horizon or outer horizon, and in extreme situation, two horizons merge) in full region of $\alpha$, when $Q^2<\frac{3\sqrt{3}}{2}$. And there are still two horizons in $1-Q^2>a>Q^2$, but only outer horizon in $Q^2>a>-\frac{4}{27}Q^6$ when $Q^2>\frac{3\sqrt{3}}{2}$.
In fact, there are no solutions for $f_\alpha(r)=0$ if $a<-\frac{1}{2}$. Let us make a detailed analysis.
%Although we can calculate the horizon by Eq~\ref{eq:horizon} when $a$ is less than $0.5$, we can substitute the horizon into the metric and find that the metric is not $0$.
%We found that the problem was in computing the horizon
%\begin{equation}
%1+\frac{r^2}{2\alpha}\left[1-\sqrt{1+4\alpha\left(\frac{2M}{r^3}-\frac{q^2}{r^4}\right)}\right]=0
%\end{equation}
It is obvious that the algebraic equation, $f_\alpha(r)=0$, can be simplified to be
\begin{equation}
\sqrt{1+4\alpha\left(\frac{2M}{r^3}-\frac{q^2}{r^4}\right)}=1+\frac{2\alpha}{r^2},\label{equsimp}
\end{equation}
where the both hand sides should not be negative. However, the right hand side may be negative when $\alpha<0$, which gives rise to an extra restriction to the range of $\alpha$. In order to find out such an extra condition, we substitute $r=r_{\rm H}^+$ into the right hand side of Eq.~(\ref{equsimp}) and thus obtain the extra constraint for $\alpha$,
\begin{equation}
1+\frac{2\alpha}{\left(M+\sqrt{M^2-q^2-\alpha}\right)^2}\geq 0,
\end{equation}
or its  dimensionless form,
\begin{equation}
2a\geq-(1+\sqrt{1-Q^2-a})^2.\label{consextr}
\end{equation}

\begin{figure}[h!]
    \centering
    \includegraphics[width=0.6\textwidth]{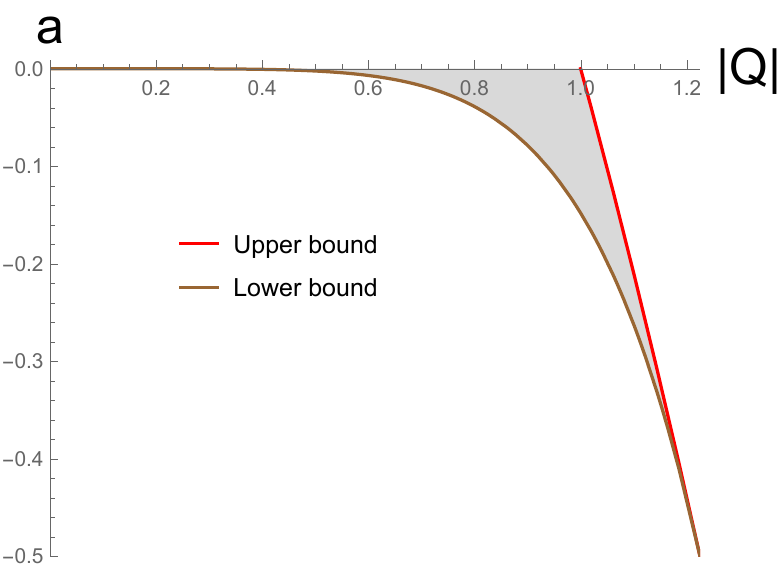}
    \caption{The horizontal axis denotes the absolute value of the rescaled charge $|Q|$ and the vertical axis the rescaled EGB parameter $a$. The shadow area determines the parameter region allowed by the set of constraints Eqs.~(\ref{constcom}), (\ref{constcom2}), and (\ref{consextr}), in which the metric function is finite everywhere.}
    \label{Fig: The region of $a$ 2}
\end{figure}

It is obvious from Eq.~(\ref{consextr}) that the existence of horizons requires the condition, $a\le 1-Q^2$, which gives the upper bound,
\begin{equation} 
a=1-Q^2,\label{redcurve}
\end{equation}
see the red curve in Fig.~\ref{Fig: The region of $a$ 2}. When $a$ takes its minimum, $a_{\rm min}=-\frac{1}{2}$, $|Q|$ takes its maximum, $|Q|_{\rm max}=\sqrt{\frac{3}{2}}$. Moreover,
Eqs.~(\ref{constcom}) and (\ref{constcom2}) determine the lower bound,
\begin{equation}
a=-\frac{4}{27}Q^6,\label{browncurve}
\end{equation}
see the brown curve in Fig.~\ref{Fig: The region of $a$ 2}. In other words, we can fix the physical region of $Q$ and $a$ by combining the constraints Eqs.~(\ref{constcom}), (\ref{constcom2}), and (\ref{consextr}), and depict it in Fig.~\ref{Fig: The region of $a$ 2}, where it is clear that $a<-\frac{1}{2}$ is beyond this physical region.

As a result, the metric function of the charged 4D EGB black hole is finite everywhere when $|Q|$ and $a$ take values within the gray area in Fig.~\ref{Fig: The region of $a$ 2}.

\subsubsection{Ricci scalar and Kretschmann scalar}
It is obvious that the spacetime of charged 4D EGB black holes has  singularity when the EGB parameter is negative, which can be seen clearly from the asymptotic form of the Ricci scalar in the limit of $r\to 0$,
\begin{equation}
\lim_{r\rightarrow 0}R\sim -\frac{4q}{r^2\sqrt{-\alpha}},\label{riccile0}
\end{equation}
and the asymptotic form of the Kretschmann scalar in the limit of $r\to 0$,
\begin{equation}
\lim_{r\rightarrow 0}K\sim -\frac{4q^2}{r^4\alpha}.\label{kretle0}
\end{equation}

\subsubsection{Radial geodesic}

We calculate the square of radial velocities of a neutral particle falling into the center of the charged 4D EGB black
hole and find that it is same as Eq.~(\ref{sqrvel4degb}) but it is associated with $\alpha<0$.
%Since in the charged case, the velocity of the particle near the origin is
%\begin{equation}
%\left(\frac{dr}{d\tau}\right)^2=E^2-1-\frac{r^2}{2\alpha}\left[1-\sqrt{1+4\alpha\left(\frac{2M}{r^3}-\frac{q^2}{r^4}\right)}\right].
%\end{equation}
%and the velocity of a free particle falling into the center of the black hole
We can see that this square is negative for a free particle ($E^2=1$) when $r\to 0$. This shows that no physical observer can reach the black hole center in a finite proper time.
%, resulting in the spacetime completeness.
%In other words, the finite metric function of the charged 4D EGB black hole leads to a complete radial geodesic, which is a part of properties of a non-singular spacetime.
Moreover, we note that the innermost radius that the particle can reach is same for the cases of $\alpha >0$, $\alpha <0$, and $\alpha \to 0$ (the vanishing case corresponds to the RN black hole) because Eq.~(\ref{innerradius}) is independent of $\alpha$.

We make a brief summary of this subsection. Whether the metric function of the charged 4D EGB black hole is complex ($\alpha > 0$) or real and finite ($\alpha < 0$) at the center of this black hole spacetime, the Ricci scalar is imaginary (Eq.~(\ref{riccige0})) or infinite (Eq.~(\ref{riccile0})), and the Kretschmann scalar is  infinite  (Eq.~(\ref{kretge0}) and Eq.~(\ref{kretle0})), while the radial geodesics are complete. This presents the peculiarity  of the curvature invariants of the charged 4D EGB black hole. We make a note that the peculiarity also appears in the complete radial geodesics, where the completeness is based on an imaginary velocity\footnote[3]{Usually the completeness is based on an infinite proper time.} which never leads to a finite proper time for a free falling particle to reach the center.

\subsection{Cosmic censorship conjecture}
The cosmic censorship conjecture requires that the singularity in a black hole spacetime should always be surrounded by an event horizon in order to prevent a naked singularity from destroying the causal structure of the spacetime.
We expect that the cosmic censorship conjecture is satisfied for the charged 4D EGB black hole.
Here we shall use the Gedanken experiment~\cite{N1} combined with the superradiation theory~\cite{N2} to verify whether the cosmic censorship conjecture is maintained or not.
The idea is to shoot a particle into the black hole and test whether the event horizon of this black hole will be destroyed.
If the event horizon is destroyed, the cosmic censorship conjecture is violated; if not, the cosmic censorship conjecture is maintained.
For the charged 4D EGB black hole with mass $M_0$ and charge $q_0$, we know from Eq.~(\ref{eq:horizon}) that the event horizon exists under the condition,
\begin{equation}
M_0^2\geq q_0^2+\alpha,
\end{equation}
where the equal sign means the case of the extreme black hole.
The black hole mass becomes $M_0+\delta M$ and the charge $q_0+\delta q$ after the black hole absorbs an incident scalar particle of frequency $\omega$ and charge $q_s$, where $\omega\ll M_0$ and $q_s\ll q_0$.
Obviously,  if the event horizon were destroyed, the mass and charge would satisfy
\begin{equation}
(M_0+\delta M)^2<(q_0+\delta q)^2+\alpha,
\end{equation}
which can be simplified when the second-order infinitesimal quantities are omitted,
\begin{equation}
\delta M<\frac{q_0^2+\alpha-M_0^2+2q_0\delta q}{2M_0}.\label{ccccond}
\end{equation}

Now let us analyze the above inequality. For the nonextreme case of  the charged 4D EGB black hole, Eq.~(\ref{ccccond})  is never satisfied and thus the cosmic censorship conjecture is maintained because the right-hand side is negative but $\delta M$ should be positive when the first-order infinitesimal quantity on the numerator is negligible.
For the case of the extreme black hole,  $M_0^2=q_0^2+\alpha$, Eq.~(\ref{ccccond}) reduces to
\begin{equation}
\delta M<\frac{q_0\delta q}{\sqrt{q_0^2+\alpha}}.
\label{ccc}
\end{equation}
Following Ref.~\cite{N2}, we know that the interaction between a static charged BH and a particle gives the relation,
\begin{equation}
\frac{\delta q}{\delta M}=\frac{q_s}{\omega},
\end{equation}
and that Eq.~(\ref{ccc}) reads
\begin{equation}
\omega<\frac{q_0}{\sqrt{q_0^2+\alpha}}q_s=\Phi_{\rm H} q_s,\label{Range}
\end{equation}
where $\Phi_{\rm H}$ is the electric potential at the event horizon.
The relationship among reflection, incidence and transmission amplitudes $\mathcal{R}, \mathcal{I}$ and $\mathcal{T}$ of the incident particle is \cite{N2}
\begin{equation}
|\mathcal{R}|^2=|\mathcal{I}|^2-\frac{\omega-\Phi_{\rm H}q_s}{\sqrt{\omega^2-\mu^2_s}}|\mathcal{T}|^2,\label{ritrel}
\end{equation}
where $\mu_s$ is particle mass.
%{\color{red}Refer to for specific proof of the above calculations.}
According to Eqs.~(\ref{Range}) and (\ref{ritrel}),  $|\mathcal{R}|^2$ is bigger than $|\mathcal{I}|^2$ in the frequency range of Eq.~(\ref{Range}), which means that the black hole will emits particles instead of absorbing, known as the superradiation phenomenon, where the mass and charge of the black hole decrease at the same time and the extreme black hole evolves into a sub-extreme configuration. This is contradictory to the precondition that no sub-extreme black holes exist. As a result, the cosmic censorship conjecture guarantees that the singularity can never be naked, which
also means that the internal structure behind the event horizon is not able to be observed directly.

\subsection{Internal structure of unreachable core regions}
\begin{figure}[h!]
    \centering
    \includegraphics[width=0.6\textwidth]{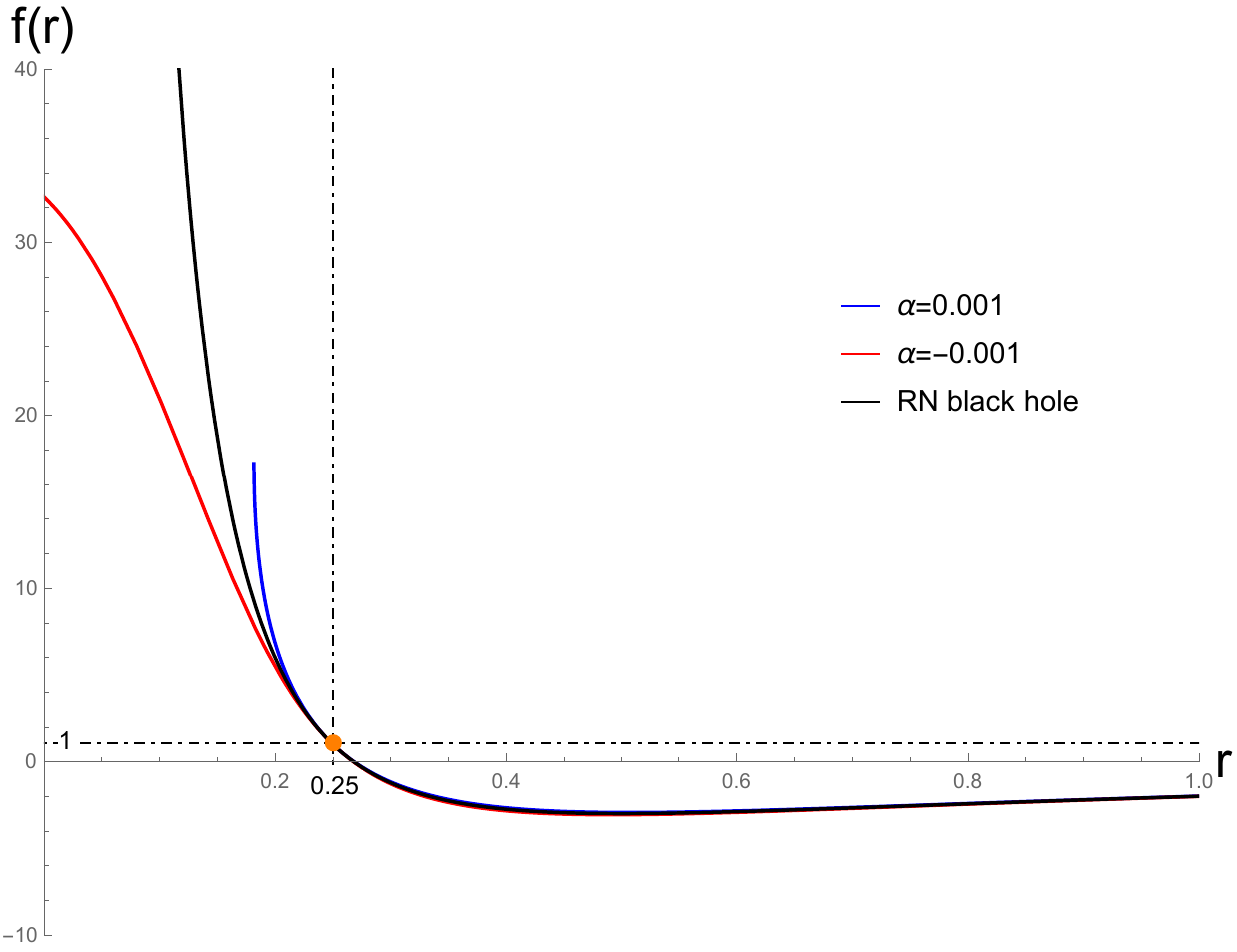}
    \caption{The horizontal axis denotes the radial coordinate and the vertical axis the metric function. The black curve represents the RN black hole, and the blue and red curves represent the charged 4D EGB black hole with the EGB parameters $0.001$ and $-0.001$, respectively, where the orange dot is the intersection of three curves. Note that the absolute value of the physically allowed $\alpha$ is very small due to $|Q|\equiv |q|/M=1/2$,  see Fig. \ref{Fig: The region of $a$ 2}.}
    \label{Fig: geodesic analysis}
\end{figure}

We have known from the discussions in subsections 2.1 and 2.2 that Eq.~(\ref{innerradius}) gives the innermost radius that a time-like particle can reach. This means that the geodesic of the particle terminates on the same sphere surface for both the complex and finite metric functions when the mass and charge are fixed. Using Eqs.~(\ref{eq:horizon}) and (\ref{innerradius}) we can check that this radius is less than the inner event horizon, i.e., $r_{\rm in}<r_{\rm H}^-$.
Thus, the sphere surface with radius $r_{\rm in}$ wraps around a region that is physically unobservable.
We can give a close relation between the core region and the metric function: On the surface of the core region the metric function equals one for the three cases --- the charged 4D EGB black holes with positive, negative, and vanishing EGB parameters (the case of  vanishing EGB parameter  corresponds to the RN black hole). In other words, as $r_{\rm in}$ is independent of the EGB parameter $\alpha$, the core region is same for the metric functions of three cases. Moreover, we notice that the difference of the metric functions is obvious inside but negligible outside the core. For instance, when we take $M=2.000$ and $|q|=1.000$, we can see in Fig. \ref{Fig: geodesic analysis} that the three metric functions intersect with one point and the differences among the three cases appear apparently inside but not outside the core region.

\begin{figure}[h!]
    \centering
    \includegraphics[width=0.6\textwidth]{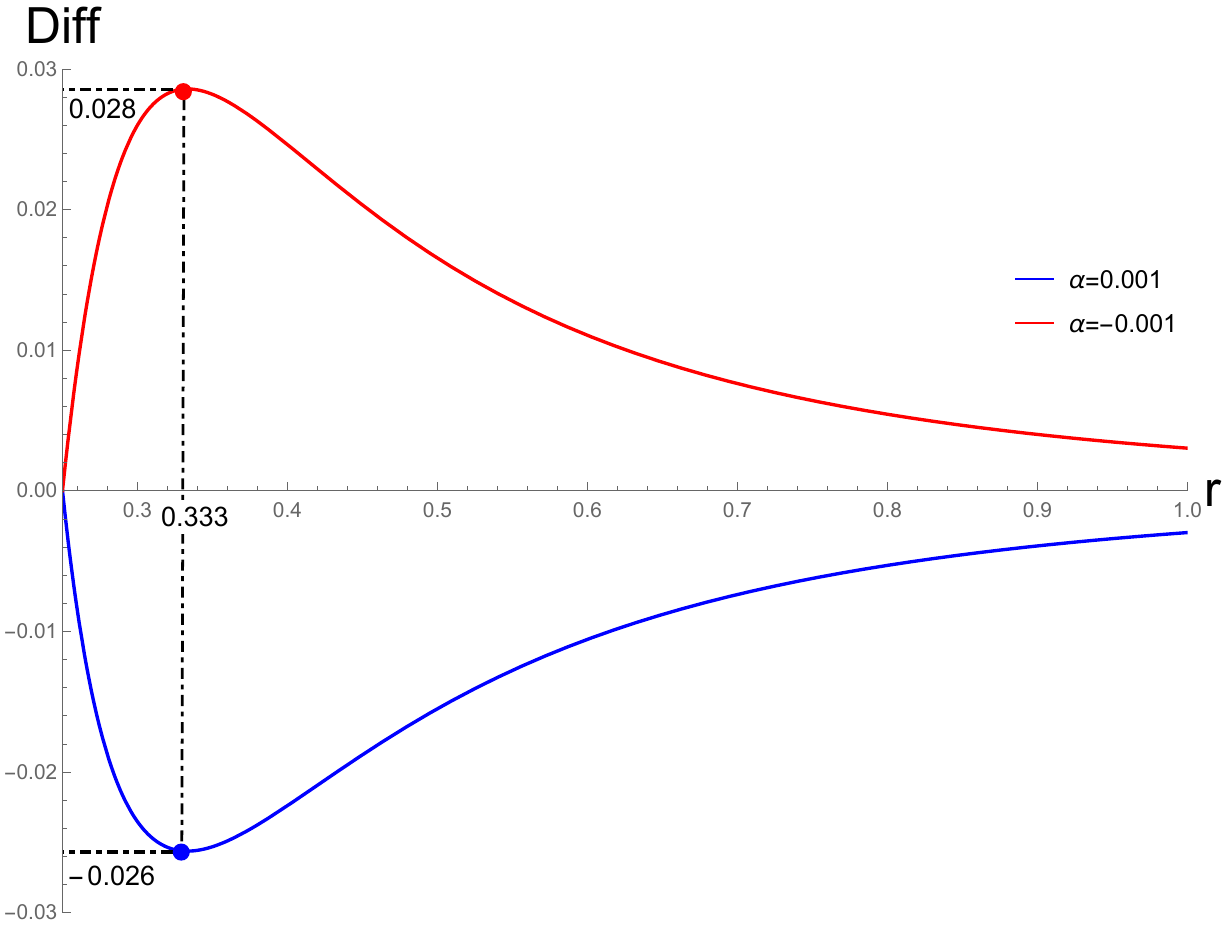}
    \caption{The horizontal axis denotes the radial coordinate and the vertical axis the relative difference of particle velocities. The blue and red curves represent the charged 4D EGB black hole with the EGB parameters $0.001$ and $-0.001$, respectively.}
    \label{Fig: diff analysis}
\end{figure}

Let us make a quantitative analysis of the relative difference between the square of radial velocities of free particles in the charged 4D EGB black hole and that in the RN black hole. The relative difference is
\begin{eqnarray}\label{relative_difference}
	Diff&\equiv&\frac{v_\alpha^2(r)-v_{\rm RN}^2(r)}{v_{\rm RN}^2(r)}=\frac{f_{\rm RN}(r)-f_\alpha(r)}{1-f_{\rm RN}(r)}\nonumber\\
	&=&\frac{r^4+4 \alpha  M r-2 \alpha  q^2-r^2 \sqrt{r^4+8 \alpha  M r-4 \alpha  q^2}}{-4 \alpha  M r+2 \alpha  q^2},
\end{eqnarray}
where $v^2$, $f_\alpha(r)$, and  $f_{\rm RN}(r)$ are given in Eq.~(\ref{velocity}), Eq.~(\ref{metric}), and Eq.~(\ref{metric_RN}), respectively, and the subscripts $\alpha$ and RN denote the charged 4D EGB black hole and the RN black hole, respectively. For the detailed derivation, see Appendix~\ref{appendix:relative_difference}. When $r\gg r_{\rm H}^+$, the absolute value of $Diff$ goes to $|\alpha| M/(2r^3)$, which is very close to zero. We plot Fig. \ref{Fig: diff analysis} by setting $M=2.000$ and $|q|=1.000$, and find that the absolute value of
the maximum relative difference of particle velocities is less than $0.030$ and this value can be reached only inside the black hole.
At the outer horizon of the black hole, $r_{\rm H}^+\approx 3.732$, $Diff\approx0.717\times 10^{-4}$ if setting $\alpha=-0.001$, and $Diff\approx-0.718\times 10^{-4}$ if setting $\alpha=0.001$, and the absolute value of $Diff$ will decrease with the increasing of the radial distance.
 Consequently, it is not realistic to distinguish the internal structures of three black holes (the charged 4D EGB black holes with positive and negative EGB parameters and the RN black hole) by the difference of particle motion.
 % and also indirectly indicates that it is difficult to distinguish the three cases by the difference in spacetime.
We thus turn to the thermodynamic and dynamic behaviors of three black holes that will represent the different internal structures of black holes.

\section{Second-order phase transition}
\label{sec:the Davies point}
Now we discuss the behaviors of Davies points associated with second-order phase transitions in the charged 4D EGB black hole.

The Davies points can be obtained from the solutions of the algebraic equation, $1/C_{q, \alpha}=0$, where $C_{q, \alpha}$, the heat capacity with a fixed $q$ and $\alpha$, is divergent at the Davies points. Here we write this heat capacity in its dimensionless form,
\begin{equation}
C_{q, \alpha}/M^2\equiv \frac{1}{M^2}\left(\frac{\partial M}{\partial T}\right)_{q, \alpha}=\frac{4\pi\,{x_{\rm H}^+}({x_{\rm H}^+}-1)({x_{\rm H}^+}^2+2a)^2}{-{x_{\rm H}^+}^4+(3{x_{\rm H}^+}^2+2a)Q^2+5{x_{\rm H}^+}^2a+2a^2},\label{hcqal}
\end{equation}
and then obtain the master equation that determines the Davies points,
\begin{equation}
-{x_{\rm H}^+}^4+(3{x_{\rm H}^+}^2+2a)Q^2+5{x_{\rm H}^+}^2a+2a^2=0,\nonumber
\end{equation}
where $Q$ and $a$ are dimensionless charge and EGB parameter, respectively, see Eq.~(\ref{rescale}). The above equation can be expressed in terms of $Q$ and $a$
when the dimensionless event horizon $x_{\rm H}^+$, Eq.~(\ref{dlhor}), is substituted,
\begin{equation}
	(4-5Q^2-7a)\sqrt{1-Q^2-a}-(7-2Q^2-4a)Q^2+2a^2-9a-4=0.\label{hcequ}
\end{equation}

\subsection{Phase transition associated with the complex metric function}
\begin{figure}[h!]
    \centering
    \includegraphics[width=0.6\textwidth]{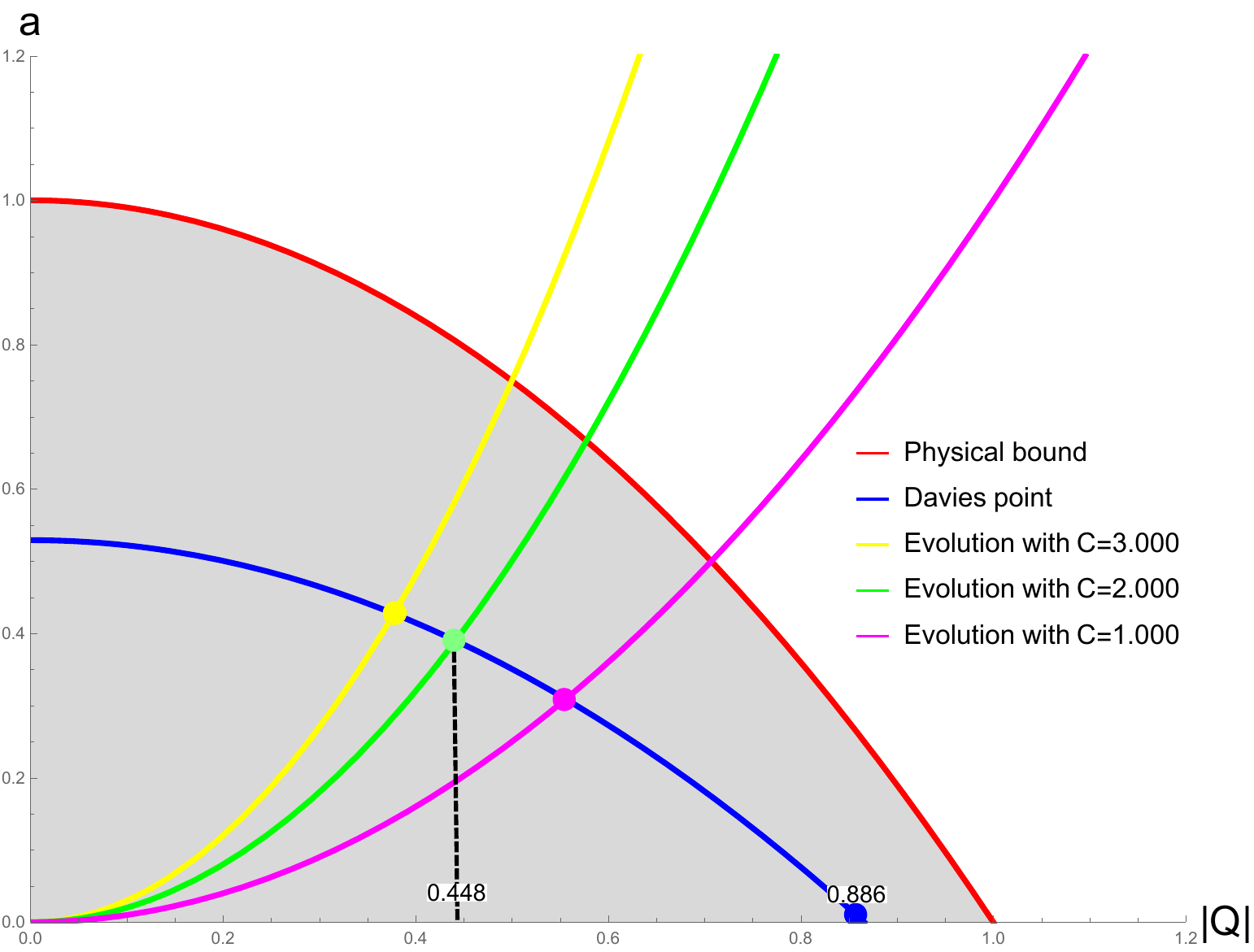}
    \caption{The behaviors of Davies points are shown on the $|Q|-a$ plane when the EGB parameter is positive for the charged 4D EGB black hole. The horizontal axis denotes the absolute value of the rescaled charge, and the vertical axis the rescaled EGB parameter. The red curve determines the boundary of the physically allowed region (in gray color) and the blue curve corresponds to the Davies points. The magenta, green, and yellow curves correspond to the evolution of the black hole with respect to mass when we take $C=1.000, 2.000, 3.000$, respectively.}
    \label{Fig: Davies point when $a>0$}
\end{figure}

For the case of the positive EGB parameter, $\alpha>0$ or $a> 0$, we plot the $|Q|-a$ graph in Fig.~\ref{Fig: Davies point when $a>0$} showing the curve of Davies points together with the curves of the physical bound and the evolution with respect to mass.

The physically allowed region of $Q$ and $a$ is determined by the fact that the charged 4D EGB black hole must have an event horizon, see Eq.~(\ref{conshor}), which, together with Eq.~(\ref{rescale}), gives the inequality, $1-Q^2-a\ge 0$. That is, the physically allowed region (in gray color) is surrounded by the red curve,\footnote[4]{It has the same formula as Eq.~(\ref{redcurve}) associated with the negative EGB parameter because the existence of horizons is same for the both cases of the positive and negative EGB parameters, see Eq.~(\ref{conshor}.)} $a=1-Q^2$, the horizontal axis, and the vertical axis for the case of $a> 0$.
The blue curve corresponds to the Davies points and is determined by Eq.~(\ref{hcequ}).

%The event horizon of the black hole exists in the region bounded by the orange line and the coordinate axis, but does not exist outside this region.

There are three parameters, $M$, $q$, and $\alpha$, that depict this black hole, and the evolution can be analyzed in principle with respect to any one of them. Considering the achievement that the Davies points have been calculated under a fixed $q$ and $\alpha$, see Eqs.~(\ref{hcqal}) and (\ref{hcequ}), we thus investigate the evolution correspondingly by fixing the two parameters but varying  the mass.
%To this end, we have to investigate whether a second-order phase transition happens or not during the evolution of the black hole.
%The evolutionary process of the black hole can be arbitrarily selected due to three variable parameters in the 4D-CEGB black hole.
%Therefore, we should limit the evolutionary process of the black hole.
%Considering that we fixed the charge and EGB parameters when calculating the heat capacity, we also fix these two parameters during the evolution of the black hole, while only changing mass of the black hole. And we'll see later that this evolution is closely related to the evolution of temperature.
%Next we should determine the corresponding curve of the evolutionary process in Fig~\ref{Fig: Davies point when $a>0$}.
Moreover, based on the definitions of the dimensionless parameters $Q$ and $a$, see Eq.~(\ref{rescale}), we obtain the following relation,
\begin{equation}
\frac{a}{Q^2}=\frac{\alpha}{q^2}=\text{const.}\equiv C,\label{defC}
\end{equation}
which gives the cluster of evolution curves,
\begin{equation}
a=CQ^2,\label{evolpos}
\end{equation} 
where $C$ is the coefficient for a fixed $q$ and $\alpha$ on the $|Q|-a$ plane.
Because $\alpha$ is positive and $|q|$ can be very small or large, $C$ can take any positive value in principle, i.e., $C\in (0, \infty)$. 
The cluster of evolution curves describes the black hole evolution with respect to mass $M$. As the heat capacity is calculated by fixing $\alpha$ and $q$, $C$ is constant for a fixed Davies point.
Obviously, this cluster of evolution curves is quadratic and passes through the origin of the $|Q|-a$ plane. For instance, we take $C=1.000, 2.000, 3.000$, which corresponds to the magenta, green, and yellow curves, respectively, in Fig.~\ref{Fig: Davies point when $a>0$}. In general, this cluster of evolution curves looks similar for any positive $C$. 
This implies that the charged 4D EGB black hole will no doubt pass through one Davies point during its evolution. In other words, one second-order phase transition must happen when the black hole evolves with respect to mass.

We note that the Davies point of the RN black hole is located at the intersection of the blue curve and the horizontal axis, corresponding to the case of $a=0$, see the blue dot in Fig.~\ref{Fig: Davies point when $a>0$}. Moreover, the charge mass ratio, $\frac{|q|}{M}$, i.e. the absolute value of the rescaled charge $|Q|$ of the charged 4D EGB black hole at the Davies points is always smaller than that of the RN black hole. This consequence is obvious when we compare the Davies point of the charged 4D EGB black hole (intersection of the blue and magenta curves, or that of the blue and green curves, or that of the blue  and yellow curves) with the Davies point of the RN black hole (intersection of the blue curve and the horizontal axis).

\subsection{Phase transition associated with the finite metric function}
\begin{figure}[h!]
\centering
\includegraphics[width=0.6\textwidth]{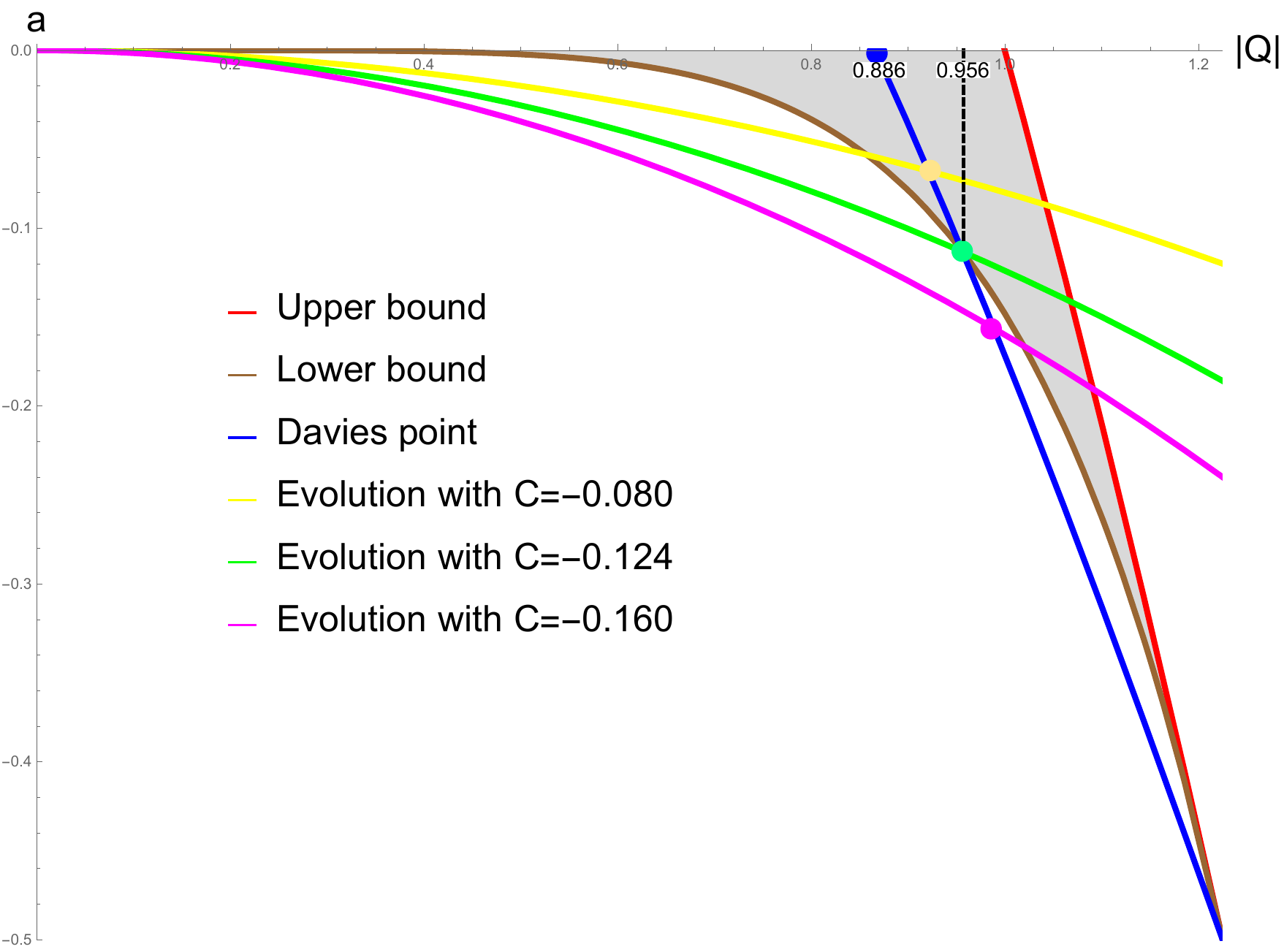}
\caption{The behaviors of Davies points are shown on the $|Q|-a$ plane when the EGB parameter is negative for the charged 4D EGB black hole. The horizontal axis denotes the absolute value of the rescaled charge, and the vertical axis the rescaled EGB parameter. The red and brown curves determine the upper and lower boundaries of the physically allowed region (in gray color), respectively. The blue curve corresponds to the Davies points. The magenta, green, and yellow curves correspond to the evolution of the black hole with respect to mass when we take $C=-0.160, -0.124, -0.080$, respectively. The brown, blue and green curves intersect at the point: $(0.956,-0.113)$.}
\label{Fig: Davies point when $a<0$}
\end{figure}

Now we turn to the case of the negative EGB parameter, $\alpha<0$ or $a< 0$. Based on Fig.~\ref{Fig: The region of $a$ 2}, we plot the $|Q|-a$ graph in Fig.~\ref{Fig: Davies point when $a<0$} showing the curve of Davies points together with the curves of the physical bound and the evolution with respect to mass.
The red and brown curves are governed by Eq. (\ref{redcurve}) and Eq.~(\ref{browncurve}), respectively, see the explanation below Eq.~(\ref{browncurve}). They determine the physical region (in gray color) in which the charged 4D EGB black hole with the finite metric function must have a horizon. The blue curve indicating the Davies points is still governed by Eq.~(\ref{hcequ}), and the cluster of evolution curves by Eq.~(\ref{evolpos}), where the former is related to a negative $a$ and the latter to a negative $C$.

%In Fig.~\ref{Fig: Davies point when $a<0$}, the shadow part represents the physical region we discussed in Sec~\ref{sec: Charged}, and the blue line is related to the Davies point.

The brown and blue curves intersect at one point: $(|Q|, a)=(0.956,-0.113)$, and the red, brown and blue curves intersect at the end of the three curves : $(|Q|, a)=(\sqrt{\frac{3}{2}},-\frac{1}{2})$, where $a$ takes its minimum. The intersection of the brown and blue curves in Fig.~\ref{Fig: Davies point when $a<0$} can be solved from the combination of their formulas, Eq.~(\ref{browncurve}) and Eq.~(\ref{hcequ}), and then the specific coefficient can be computed by its definition, Eq.~(\ref{defC}), $C=-\frac{0.113}{0.956^2}=-0.124$. As a result, we can determine the formula governing the green curve, $a=-0.124Q^2$. We can see that only the interval of the blue curve between $(|Q|, a)=(0.886, 0)$ and $(|Q|, a)=(0.956, -0.113)$ is located inside the physical region (in gray color), i.e., only the Davies points on this interval are allowed physically, where the physical permission means the existence of horizons for the 4D EGB black hole. 

In Fig.~\ref{Fig: Davies point when $a<0$}
the range of $C$ is $C\in (-\frac{1}{3}, 0)$, where the intersection of the red, brown and blue curves, $(|Q|, a)=(\sqrt{\frac{3}{2}},-\frac{1}{2})$, gives the end point of the three curves, see Eq.~(\ref{defC}).
The green dot divides the blue curve representing the Davies points into two parts, with the upper part, $C\in [-0.124, 0)$, located inside the physical region and the lower part, $C\in (-\frac{1}{3}, -0.124)$, outside the physical region.
If the cluster of evolution curves, i.e. $a=CQ^2$, goes through the upper part of the blue curve and intersects with the blue one, e.g. the yellow curve, the intersection is located inside the physical region; if it goes through the lower part of the blue curve and intersects with the blue one, e.g. the magenta curve, the intersection is located outside the physical region.
Therefore, the black holes associated with $C\in [-0.124, 0)$ will undergo a phase transition during the evolution, but those associated with $C\in (-\frac{1}{3}, -0.124)$ will not.
We can say, the green dot associated with $C=-0.124$ is the split point of whether the black hole has a phase transition or not.

%As can be seen from Fig~\ref{Fig: Davies point when $a<0$}, when the evolution curve passes through the lowest point, $c=-0.333$, which corresponds to the minimum value of $c$.
%Notice that the curve corresponding to the orange line in the figure is $a=-0.124Q^2$, so the black hole will be restricted in the physical region and cannot reach the Davies point when $-0.333<c<-0.124$.
It is worth mentioning the end point, $(|Q|, a)=(\sqrt{\frac{3}{2}},-\frac{1}{2})$, which corresponds to the coefficient $C=-\frac{1}{3}$.
As discussed in subsection~\ref{sec2.2}, the charged 4D EGB black hole at this point stays in its extreme configuration with only one horizon.
From Fig.~\ref{Fig: Davies point when $a<0$}, we can see that the end point is a Davies point, but the black hole in the extreme configuration cannot evolve to the other non-extreme configurations.
For the same reason, the other non-extreme configurations cannot evolve to the extreme configuration, either. This consequence is obvious because the blue curve between $(|Q|, a)=(0.956,-0.113)$ and $(|Q|, a)=(\sqrt{\frac{3}{2}},-\frac{1}{2})$, i.e., $C\in (-\frac{1}{3}, -0.124)$, is beyond the physical region.

Let us make a comparison between the case of the positive EGB parameter (depicted by Fig.~\ref{Fig: Davies point when $a>0$}) and the case of the negative EGB parameter (depicted by Fig.~\ref{Fig: Davies point when $a<0$}). For the former, we can see in Fig.~\ref{Fig: Davies point when $a>0$} that the whole blue curve is located inside the physical region (in gray color) and that the cluster of evolution curves, $a=CQ^2$, $C\in (0, \infty)$, intersects with the blue one for any positive $C$. In other words, all the Davies points are allowed physically.
For the latter, we can see in Fig.~\ref{Fig: Davies point when $a<0$} that only the upper part between $(|Q|, a)=(0.886, 0)$ and $(|Q|, a)=(0.956, -0.113)$ is located inside the physical region (in gray color) and that the cluster of evolution curves, $a=CQ^2$, intersects with the blue one in the physical region  when $C\in [-0.124, 0)$. In other words, only one part of the Davies points is allowed physically.
This is the crucial difference between the two cases.
In addition, the charge mass ratio $\frac{|q|}{M}$ of the charged 4D EGB black hole at the Davies points of the physical region is always smaller than that of the RN black hole for the former case (Fig.~\ref{Fig: Davies point when $a>0$}), but it is always larger than that of the RN black hole for the latter case (Fig.~\ref{Fig: Davies point when $a<0$}).

%\subsection{Summary of phase transitions associated with the singular and non-singular metrics}
%The charged 4D EGB black hole can always reach one second-order phase transition point during its evolution with respect to mass if the EGB parameter is positive. However, if the EGB parameter is negative, the situation is completely different. That is to say, the black hole can undergo one second-order phase transition if the coefficient satisfies the condition, $-0.124\le \frac{\alpha}{q^2}< 0$, but it cannot if $-\frac{1}{3}\le \frac{\alpha}{q^2}<-0.124$.

\subsection{Heat capacity and temperature at a Davies point associated with the complex and finite metric functions}
Let us discuss the behaviors of the heat capacity and temperature when a charged 4D EGB black hole evolves through a Davies point.
%\begin{equation}
%\frac{1}{C_q}=\left(\frac{\partial T}{\partial M}\right)_q=0
%\end{equation}
Based on the definition of the heat capacity with a fixed $q$ and $\alpha$, Eq.~(\ref{hcqal}), we derive the temperature during the evolution,
\begin{equation}
T=\int_{\ell} dT=\int_{\ell}\left(\frac{\partial T}{\partial M}\right)_{q, \alpha} dM=\int_{\ell} \frac{dM}{C_{q, \alpha}},
\end{equation}
where ``$\ell$" means the evolutionary path of charged 4D EGB black holes. We can conclude that a Davies point corresponds to a saddle point of the curve of temperature versus mass. As to whether the saddle point corresponds to a maximum or a minimum, we have to analyze by considering a positive EGB parameter and a negative one separately.

\begin{figure}[h!]
    \centering
    \includegraphics[width=0.6\textwidth]{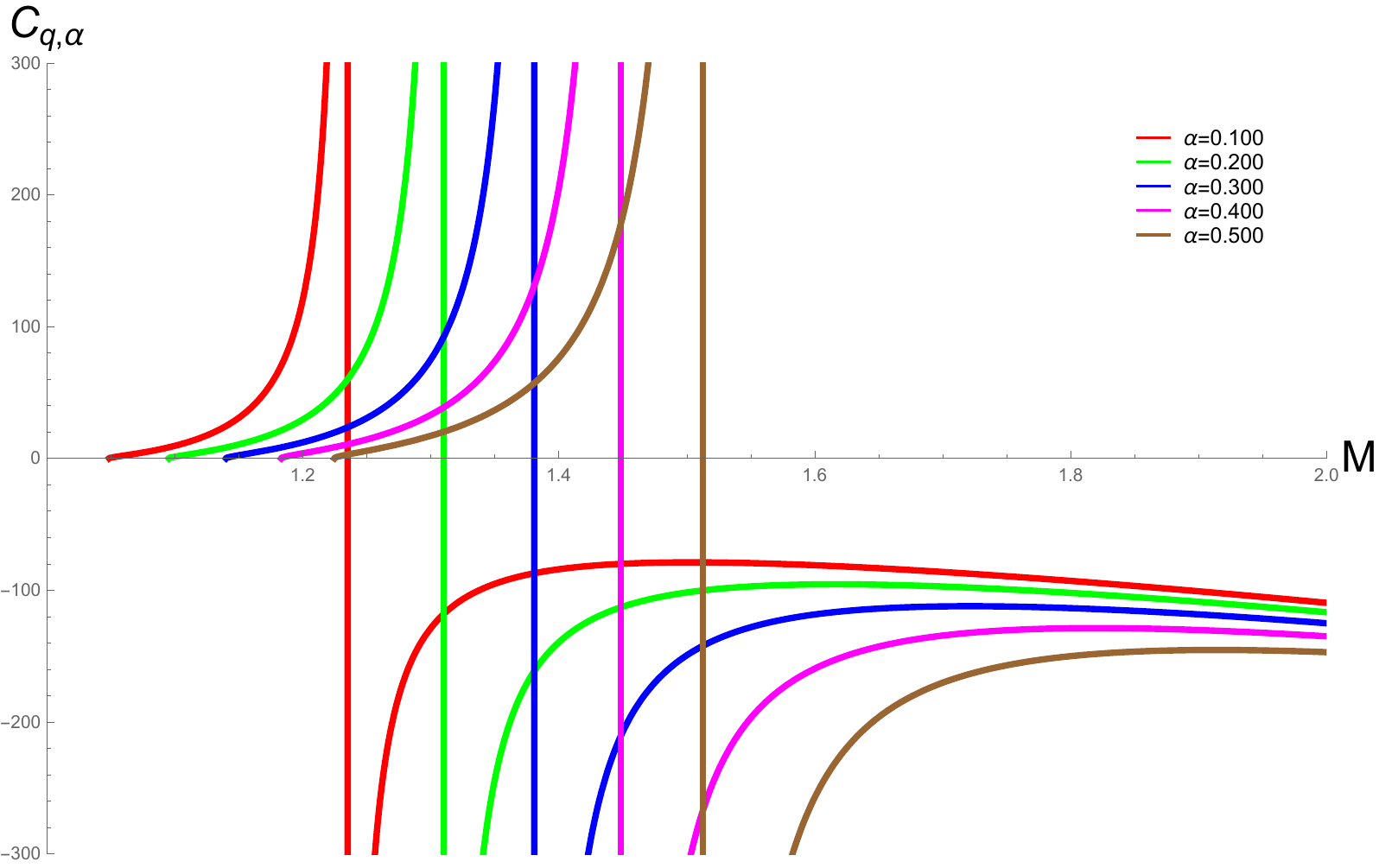}
    \caption{The relationship between the heat capacity and mass when the EGB parameter is positive for the charged 4D EGB black hole. The EGB parameter $\alpha$ is set to be $\alpha=0.100, 0.200, \dots, 0.500$, and the charge $q=1.000$. One vertical line corresponds to one Davies point for each value of $\alpha$.}
    \label{Fig: Capacity when $a>0$}
\end{figure}

%\subsubsection{Behaviors of heat capacity and temperature at a Davies point associated with the singular metric}
At first, we examine the relationship between the heat capacity and mass of the black hole when $\alpha>0$ or $a>0$. We plot the $M- C_{q, \alpha}$ graph in Fig.~\ref{Fig: Capacity when $a>0$} by using Eq.~(\ref{hcqal}) and Eq.~(\ref{dlhor}) and replacing $Q$ and $a$ by $M$, $q$, and $\alpha$. For each of $\alpha=0.100, 0.200, \dots, 0.500$ with the fixed $q=1.000$, we can see that the heat capacity changes from positive to negative when the black hole mass increases and passes through the corresponding Davies point, indicating that this Davies point is the only saddle point with the maximal temperature.

\begin{figure}[h!]
    \centering
    \includegraphics[width=0.6\textwidth]{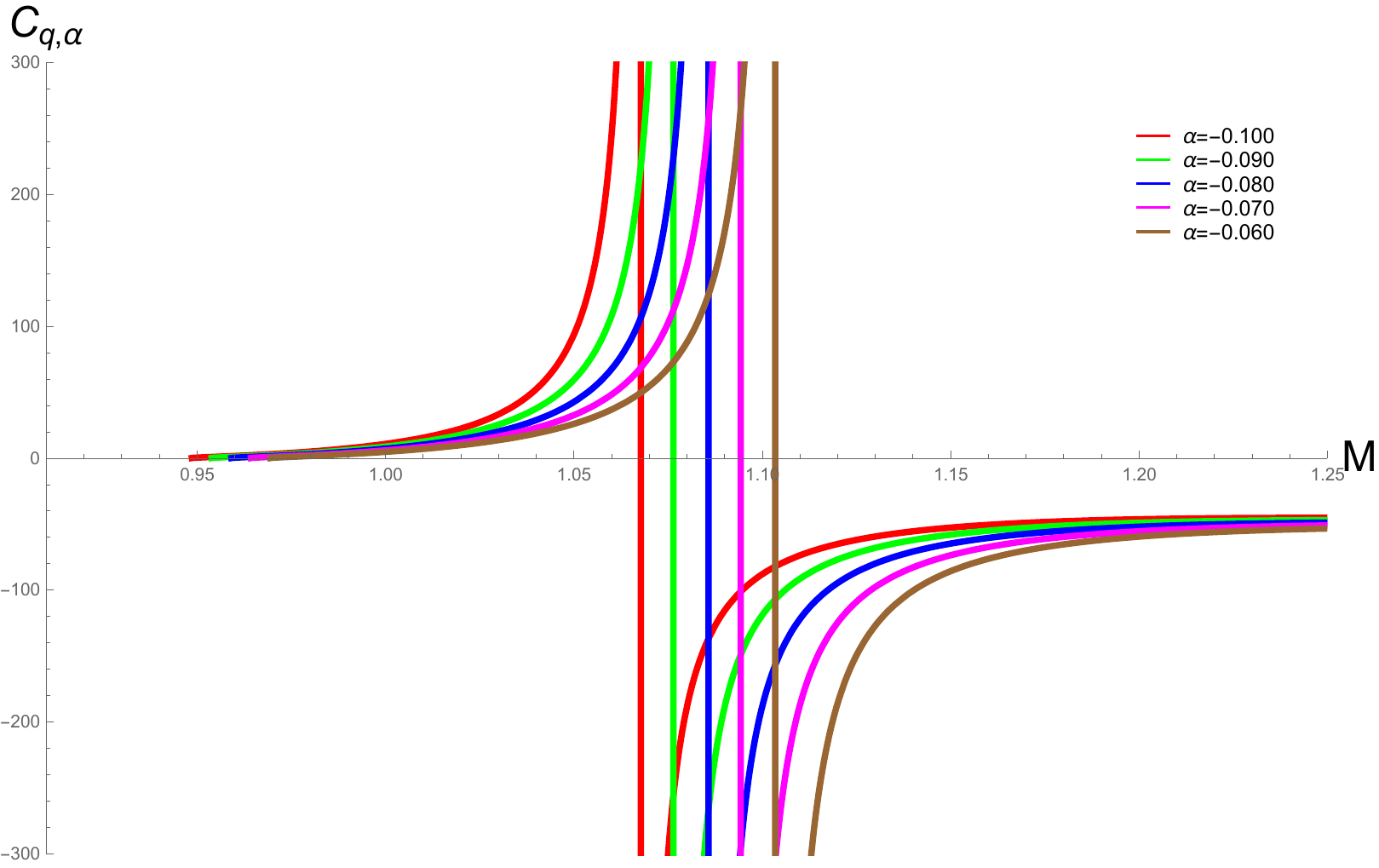}
    \caption{The relationship between the heat capacity and mass when the EGB parameter is negative and located in the range of $-0.124\le \frac{\alpha}{q^2}<0$ for the charged 4D EGB black hole. The EGB parameter $\alpha$ is set to be $\alpha=-0.100, -0.090, \dots, -0.060$, and the charge $q=1.000$. One vertical line corresponds to one Davies point for each value of $\alpha$.}
    \label{Fig: Capacity when $a<0$}
\end{figure}

\begin{figure}[h!]
    \centering
    \includegraphics[width=0.6\textwidth]{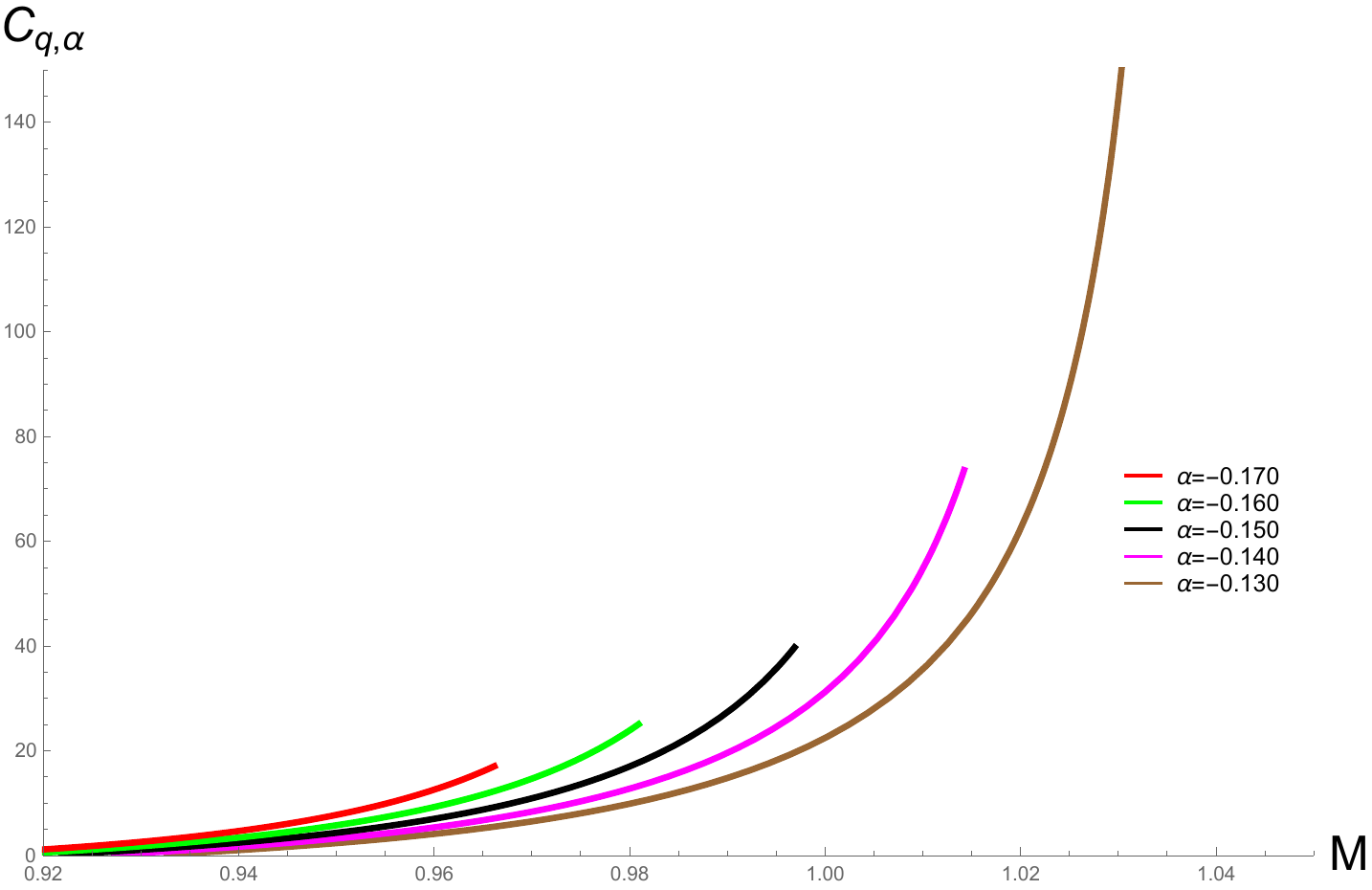}
    \caption{The relationship between the heat capacity and mass when the EGB parameter is negative and located in the range of $-\frac{1}{3}<\frac{\alpha}{q^2}<-0.124$ for the charged 4D EGB black hole. The EGB parameter $\alpha$ is set to be $\alpha=-0.170, -0.160, \dots, -0.130$, and the charge $q=1.000$.
    The black hole does not pass through any Davies points during its evolution and its temperature increases monotonously with the increasing of mass.}
    \label{Fig: Capacity without Davies when $a<0$}
\end{figure}

Then we turn to investigate the relationship between the heat capacity and mass when $\alpha<0$ or $a<0$, and classify this case into two subcases which correspond to  $-0.124\le \frac{\alpha}{q^2}<0$ and  $-\frac{1}{3}<\frac{\alpha}{q^2}<-0.124$, respectively.
As done in the above case, we plot the $M- C_{q, \alpha}$ graphs in Fig.~\ref{Fig: Capacity when $a<0$} when $\alpha$ is set to be $-0.100, -0.090, \dots, -0.060$ and $q=1.000$ for the first subcase, and in Fig.~\ref{Fig: Capacity without Davies when $a<0$} when $\alpha$ is set to be $-0.170, -0.160, \dots, -0.130$ and $q=1.000$ for the second subcase.
From Fig.~\ref{Fig: Capacity when $a<0$}, we can see that the Davies points exist in the process of evolution of the black hole, and that the heat capacity still goes from positive to negative when the mass increases and passes through the corresponding Davies point, indicating that this Davies point is the only saddle point with the maximal temperature.
From Fig.~\ref{Fig: Capacity without Davies when $a<0$}, we can see, however, that the black hole does not pass through any Davies points during its evolution, and that the temperature increases monotonously with the increasing of mass. Note that for the extreme configuration, $\frac{\alpha}{q^2}=-\frac{1}{3}$, it is isolated by the evolution although it is a Davies point as explained in subsection 3.2.

%\subsection{Summary of heat capacity and temperature at a Davies point associated with the singular and non-singular metrics}
%When the EGB parameter is positive or negative in the range of $-0.124\le \frac{\alpha}{q^2}<0$, the charged 4D EGB black hole will pass through one Davies point during its evolution which corresponds to the only saddle point with the maximal temperature. When the EGB parameter is negative in the range of $-\frac{1}{3}<\frac{\alpha}{q^2}<-0.124$, the charged 4D EGB black hole will, however, not pass through any Davies points and its temperature is going up monotonously if its mass is increasing.

\section{Quasinormal modes in the eikonal limit}
\label{sec:QNM}
According to the light ring/quasinormal mode correspondence~\cite{P10},the quasinormal mode frequencies of a test scalar fields perturbation in the eikonal limit are of the following form for a static and spherically symmetric black hole,
\begin{equation}
\omega=\Omega_c l-i(n+1/2)|\lambda|,
\end{equation}
where the angular velocity $\Omega_c$ and the Lyapunov index $\lambda$ can be computed by
\begin{equation}
\Omega_c=\frac{\sqrt{f_c}}{r_c},\qquad \lambda =\sqrt{\frac{f_{c}(2f_{c}-r_{c}^2 f''_{c})}{2 r_{c}^2}}.\label{omelam}
\end{equation}
Here $l$ is the multiple number, where the eikonal limit means $l\gg 1$; $n$ is the overtone number;
$f_{c}\equiv f(r_{c})$ and $f''_{c}\equiv f''(r)|_{r=r_{c}}$, where the prime means the derivative with respect to $r$ and $r_c$, the radius of a photon sphere, satisfies the following equation,
\begin{equation}
2f(r_c)-rf'(r_c)=0.\label{eqpsr}
\end{equation}

When substituting the shape function Eq.~(\ref{metric}) of the charged 4D EGB black hole into Eq.~(\ref{eqpsr}), we obtain the relationship between $r_c$ and the three parameters of this black hole,
\begin{equation}
r_c^4-9M^2r_c^2+4M(3q^2+\alpha)r_c-4q^2(q^2+\alpha)=0,
\end{equation}
and we also derive the angular velocity $\Omega_c$ and the Lyapunov index $\lambda$ by substituting Eq.~(\ref{metric}) into Eq.~(\ref{omelam}),
\begin{subequations}
\begin{equation}
\Omega_c^2=\frac{f_c}{r_c^2}=\frac{1}{r_c^2}+\frac{1-A^{1/2}(r_c)}{2\alpha},
\end{equation}
\begin{equation}
\lambda^2=\frac{f_c(2f_c-r_c^2f''_c)}{2r_c^2}=\frac{f(r_c)}{r_c^8\,A^{3/2}(r_c)}\left(r_c^6\,A^{3/2}(r_c)-2r_c^4q^2-36r_c^2M^2\alpha+32r_cMq^2\alpha-8q^4\alpha\right),
\end{equation}
\end{subequations}
where
\begin{equation}
A(r_c)\equiv 1+4\alpha\left(\frac{2M}{r_c^3}-\frac{q^2}{r_c^4}\right).
\end{equation}
By using the rescaled quantities, $Q$, $a$, and $x_c$, see Eqs.~(\ref{rescale}) and (\ref{rescalex}), we can rewrite the above relations in a dimensionless form,
\begin{equation}
x_c^4-9x_c^2+4(3Q^2+a)x_c-4Q^2(Q^2+a)=0,\label{posxc}
\end{equation}
\begin{equation}
\Omega_c^2M^2=\frac{1}{x_c^2}+\frac{1-A^{1/2}(x_c)}{2a},\label{posomega}
\end{equation}
\begin{equation}
\lambda^2M^2=\frac{f(x_c)}{x_c^8\,A^{3/2}(x_c)}\left(x_c^6\,A^{3/2}(x_c)-2x_c^4Q^2-36x_c^2a+32x_cQ^2a-8Q^4a\right),\label{posxlamdda}
\end{equation}
where $f(x_c)$ and $A(x_c)$ are defined by
\begin{equation}
f(x_c)\equiv 1+\frac{x_c^2}{2a}\left[1-\sqrt{1+4a\left(\frac{2}{x_c^3}-\frac{Q^2}{x_c^4}\right)}\right],\label{posAc}
\end{equation}
\begin{equation}
A(x_c)\equiv 1+4a\left(\frac{2}{x_c^3}-\frac{Q^2}{x_c^4}\right).\label{posfc}
\end{equation}

\subsection{Quasinormal modes in the eikonal limit associated with the complex metric function}

\begin{figure}[h!]
    \centering
    \includegraphics[width=0.6\textwidth]{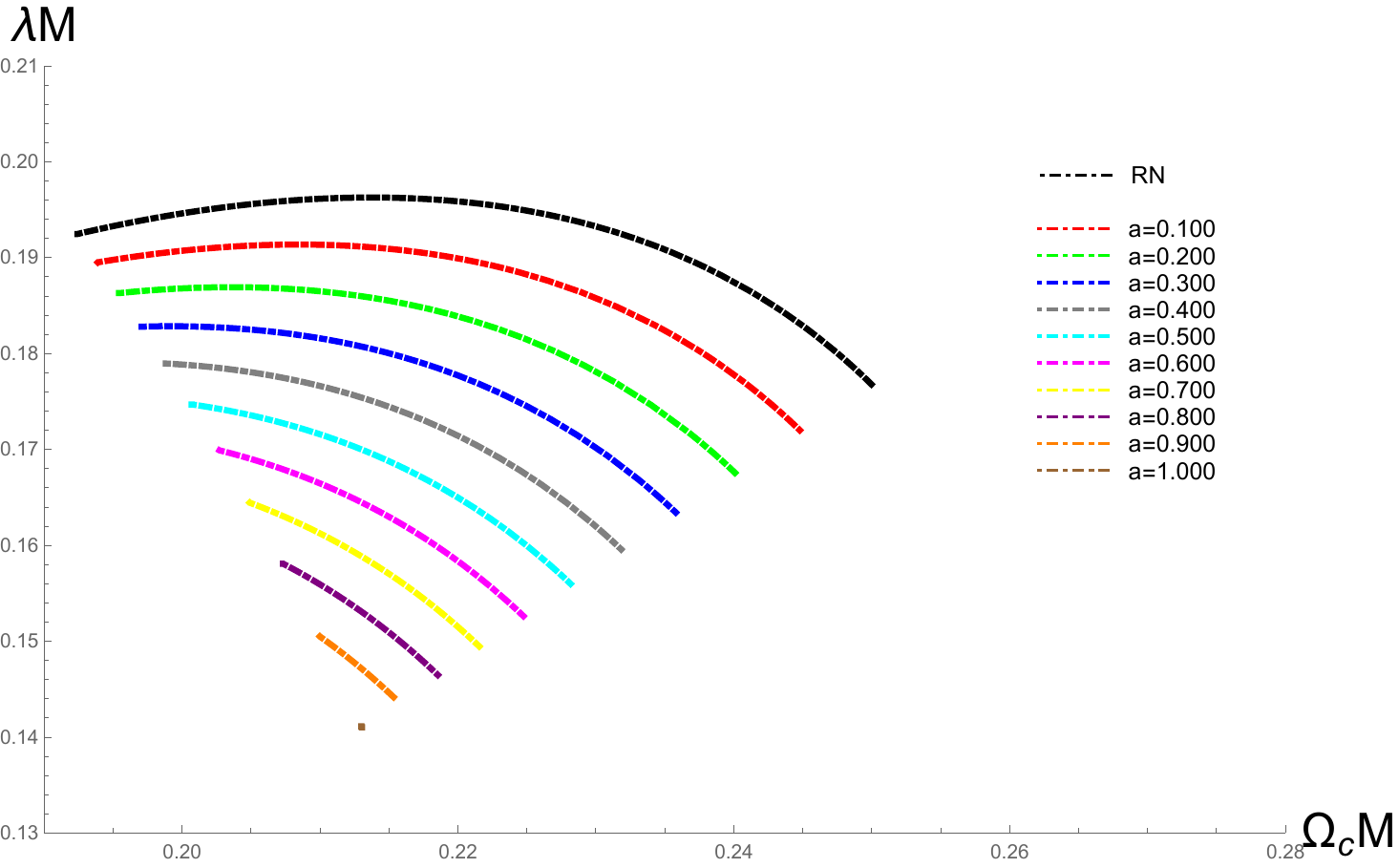}
    \caption{The relationship between the angular velocity and Lyapunov index for the positive EGB parameter. The horizontal axis denotes the dimensionless angular velocity, and the vertical one the dimensionless Lyapunov index. The black curve corresponds to the  RN black hole and the other curves correspond to the charged 4D EGB black hole with different values of the positive EGB parameter.}
    \label{Fig: eiknol QNM when a>0}
\end{figure}

Now we plot the $\Omega_c M-\lambda M$ graph for the positive EGB parameter ($\alpha>0$ or $a>0$) in Fig.~\ref{Fig: eiknol QNM when a>0} in terms of Eqs.~(\ref{posxc})-(\ref{posfc}), where the parameter is set to be $a=0.100, 0.200, \dots, 0.900$, and $a=0$ corresponding to the RN black hole is attached for comparison.
Note that the range of $Q$ that is physically allowed in the figure depends on the range of $a$, that is, the ranges of $Q$ and $a$ should ensure the existence of a horizon, i.e., to satisfy Eq.~(\ref{dlhor}).

We can see from Fig.~\ref{Fig: eiknol QNM when a>0} that 
%the angular velocity of the photon sphere of the charged 4D EGB black hole gradually decreases with the increasing of the EGB parameter, and that 
the ranges of the angular velocity for different values of the parameter $a$ are restricted within the range of the angular velocity of the RN black hole.
For instance, see the red curve ($a=0.100$), its range of the horizontal coordinate is completely covered by the range of the horizontal coordinate of the black curve ($a=0$). In other words, we cannot distinguish the 4D EGB black hole from the RN black hole when we observe the angular velocity only.
Note that the angular velocity of the photon sphere corresponds to the oscillating frequency in quasinormal modes. That is to say, we cannot distinguish the charged 4D EGB black hole from the RN black hole just by means of the oscillating frequency of quasinormal modes.
Fortunately, the range of the Lyapunov index for the charged 4D EGB black hole cannot be covered completely by that of the RN black hole. For instance, the minimum value of $\lambda M$ in the 4D EBG black hole is $0.141$ which appears when $a=1.000$ and $Q=0$,\footnote[5]{When $a$ is taken to be one, $Q$ must be zero in order to ensure the existence of a horizon. That is, the charged 4D EGB black hole becomes its uncharged case if $a=1.000$, see Eq.~(\ref{dlhor}).} 
while the range of $\lambda M$ in the RN black hole is $0.177\le \lambda M\le 0.196$.
This implies that the Lyapunov index of the charged 4D EGB black hole can be smaller than the minimum Lyapunov index of the RN black hole. Therefore, we have the possibility to distinguish the charged 4D EGB black hole from the RN black hole when we observe the Lyapunov index.

It is known that the imaginary part of quasinormal mode frequencies is inversely proportional to the damping time of a test scalar field perturbation,
\begin{equation}
 \tau\propto\frac{1}{|\omega_{\rm I}|},
 \end{equation}
which means the time past for the amplitude of a test scalar field perturbation to decay to $e^{-1}$ of its original value. Thus, the damping time of the uncharged 4D EGB black hole will be longer than the maximum damping time of the RN black hole if $a$ is appropriately chosen,\footnote[6]{This consequence is also valid for the charged 4D EGB black hole if $Q$ and $a$ are appropriately chosen in accordance with Eq.~(\ref{dlhor}).} and the maximum relative deviation between the damping time of 4D EGB black holes and that of the RN black hole equals
\begin{equation}
\frac{\Delta \tau_1}{\tau_1}=\frac{\frac{1}{0.141}-\frac{1}{0.177}}{\frac{1}{0.177}}\approx25.5\%.
\end{equation}

\subsection{Quasinormal modes in the eikonal limit and shadow radii associated with the finite metric function}

\begin{figure}[h!]
    \centering
    \includegraphics[width=0.6\textwidth]{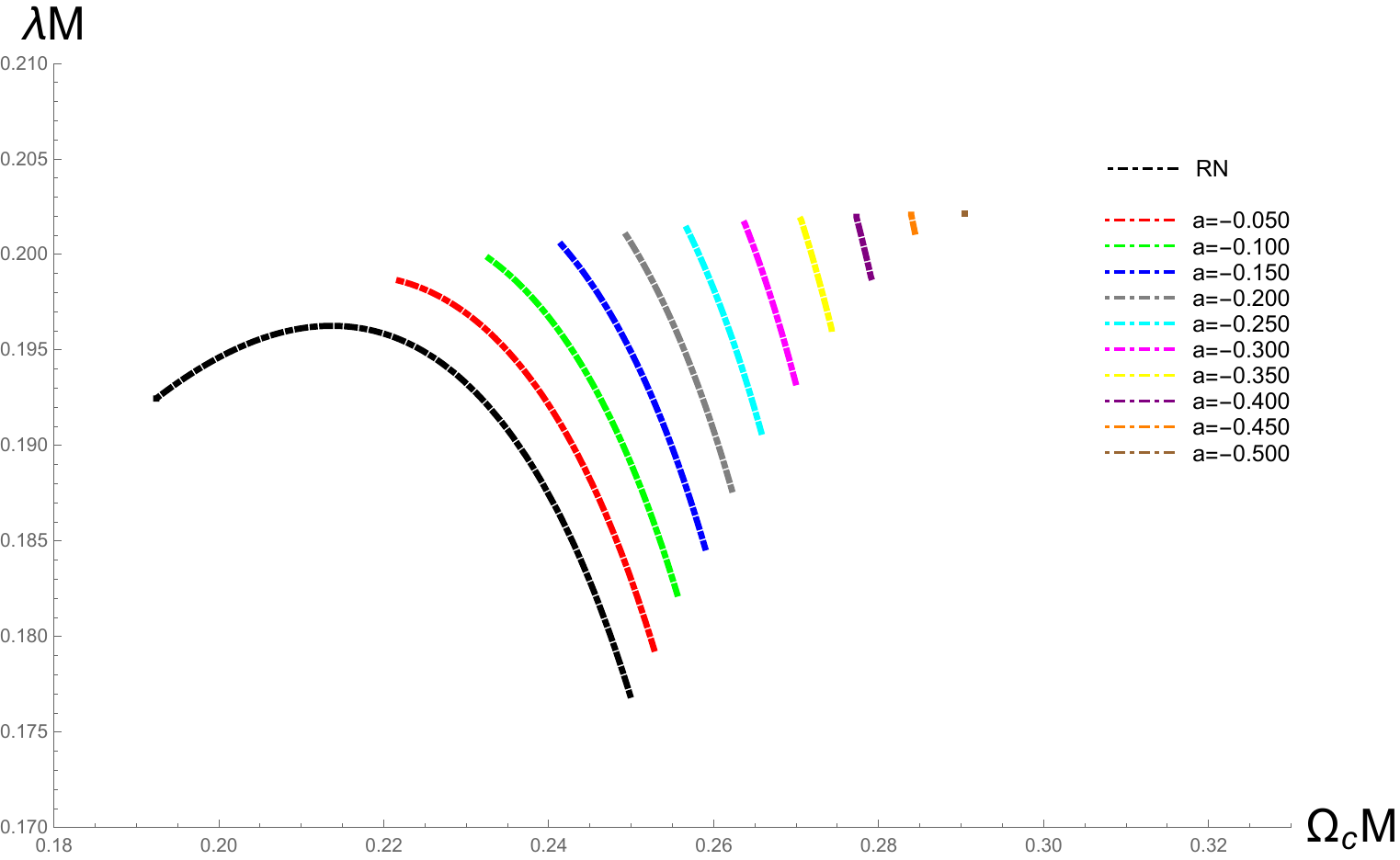}
    \caption{The relationship between the angular velocity and Lyapunov index for the negative EGB parameter. The horizontal axis denotes the dimensionless angular velocity, and the vertical one the dimensionless Lyapunov index. The black curve corresponds to the RN black hole and the other curves correspond to the charged 4D EGB black hole with different values of the negative EGB parameter.}
    \label{Fig: eiknol QNM when a<0}
\end{figure}

Now we plot the $\Omega_cM-\lambda M$ graph for the negative EGB parameter ($\alpha<0$ or $a<0$) in Fig.~\ref{Fig: eiknol QNM when a<0} in terms of Eqs.~(\ref{posxc})-(\ref{posfc}), where the parameter is set to be $a=-0.050, -0.100, \dots, -0.450$, and $a=0$ corresponding to the RN black hole is attached for comparison.
Note that the range of $Q$ that is physically allowed in the figure depends on the range of $a$, that is,
the ranges of $Q$ and $a$ should coincide with the constraints Eqs.~(\ref{constcom}), (\ref{constcom2}), and (\ref{consextr}), i.e., they have to be located in the physical region depicted in Fig.~\ref{Fig: The region of $a$ 2}.

We can see from Fig.~\ref{Fig: eiknol QNM when a<0} that both the range of the angular velocity and that of the Lyapunov index for the charged 4D EGB black hole cannot be covered completely by those of the RN black hole.
For the angular velocity, the maximum value of $\Omega_cM$ for the charged 4D EGB black hole is $0.291$ at $a=-0.500$, while the range for the RN black hole is $0.192\le \Omega_cM\le 0.250$. This means that the angular velocity of the 4D EGB black hole is larger than the maximum value of the RN black hole. Therefore, we have the possibility to distinguish the two types of black holes when we observe the angular velocity.
For the Lyapunov exponent, the maximum value of the charged 4D EGB black hole can be up to $0.202$ when $a=-0.500$, while the range of $\lambda M$ for the RN black hole is $0.177\le\lambda M\le0.196$. This implies that the Lyapunov exponent of the charged 4D EGB black hole is larger than the maximum of the RN black hole, which suggests that we can also have the possibility to distinguish the two types of black holes when we observe the Lyapunov exponent.

As a result, the damping time of the charged 4D EGB black hole can be shorter than the minimum damping time of the RN black hole, and the maximum relative deviation between them is
\begin{equation}
\frac{\Delta \tau_2}{\tau_2}=\frac{\frac{1}{0.202}-\frac{1}{0.196}}{\frac{1}{0.196}}\approx-3.0\%,
\end{equation}
where the minus sign means that the damping time of the charged 4D EGB black hole is smaller than the minimum damping time of the RN black hole.
Moreover, the oscillating frequency of the charged 4D EGB black hole can be higher than the maximum oscillating frequency of the RN black hole, and the maximum relative deviation between them is
\begin{equation}
 \frac{\Delta \omega_{\rm R2}}{\omega_{\rm R2}}=\frac{0.291-0.250}{0.250}=16.4\%.
\end{equation}

Alternatively, we may distinguish the charged 4D EGB black hole from the RN black hole by using their shadow radii. The shadow radius $R_{\rm sh}$ observed at infinity outside a black hole is closely related~\cite{P12} to the angular velocity of the photon sphere,
\begin{equation}
R_{\rm sh}=\frac{r_c}{\sqrt{f(r_c)}}=\Omega_c^{-1}.
\end{equation}
Therefore, the shadow radius of the charged 4D EGB black hole can be smaller than the minimum shadow radius of the RN black hole when the EGB parameter is negative, and the maximum relative deviation between them is
\begin{equation}
\frac{\Delta R_{\rm sh}}{R_{\rm sh}}=\frac{\frac{1}{0.291}-\frac{1}{0.250}}{\frac{1}{0.250}}\approx-14.1\%,
\end{equation}
where the minus sign has the similar meaning to that mentioned above.
As a result, the shadow radius of the charged 4D EGB black hole is smaller than the smallest shadow radius of the RN black hole in the case of the finite metric function, which does not happen in the case of the complex metric function. 

Comparing the case of the complex metric function (positive EGB parameter) with the case of the finite metric function (negative EGB parameter), we find their dynamic difference in the angular velocity and Lyapunov exponent or in the oscillating frequency and damping time. This may give the possibility to shed some light on the interior of black holes (depicted partially by metric functions) from the exterior (shown partially by dynamic properties).

\section{Conclusion}

\label{sec:conclusions}
In this paper, our main work is to connect an undetectable property of black holes --- the metric singularity at a black hole center to a probably observable phenomenon or variable, such as the second-order phase transition of black holes, the oscillating frequency and damping time of perturbation of a test scalar field, and the shadow radius of black holes.
%we reveal a contradiction between the completeness of spacetime and {\color{red}the requirement of the Einstein-Gauss-Bonnet parameter in the uncharged case}. Its significance lies in the finding that the incomplete regions of the 4D EGB black holes depend on the ranges of the EGB parameter, which expands the concept of spacetime completeness from only the coordinate dependence to the coordinate-and-parameter dependence.
For the charged 4D EGB black hole, if the EGB parameter is positive, the metric is singular at the center; if the EGB parameter is negative and located in the physically allowed region shown in Fig.~\ref{Fig: The region of $a$ 2}, the metric is non-singular.
%In order to judge the Nature prefers a singular or non-singular metric, we take the Reissner-Nordstr\"om black hole of Einstein's general relativity as an object of reference, and
We try to distinguish the charged 4D EGB black hole with or without the metric function singularity from the Reissner-Nordstr\"om black hole by investigating the behaviors of the Davies point and calculating the quasinormal modes in the eikonal limit. We summarize the following differences between the characteristic variables of thermodynamics and dynamics in the charged 4D EGB black hole and the same variables in the Reissner-Nordstr\"om black hole.% and expect these differences to be observed in future experiments.

\begin{itemize}
    \item %[1)]
    When the EGB parameter is positive or negative in the range of $-0.124\le \frac{\alpha}{q^2}<0$, the charged 4D EGB black hole will pass through one Davies point in the evolution with respect to mass, which indicates that there exists one unique saddle point with the maximum temperature. However, if the EGB parameter is negative in the range of $-\frac{1}{3}< \frac{\alpha}{q^2}<-0.124$, the black hole will not pass through a Davies point in the evolution and its temperature will not have an extremum, either. Compared with the second-order phase transition point of the RN black hole, the charge mass ratio $\frac{|q|}{M}$ at the Davies point is smaller than that of the RN black hole for the positive EGB parameter, while it is larger than that of the RN black hole for the negative EGB parameter of the range $-0.124\le \frac{\alpha}{q^2}<0$.
    \item%[2)]
    When the EGB parameter is positive, the oscillating frequency of quasinormal modes of a test scalar field perturbation in the eikonal limit will not exceed the range of oscillating frequencies of the RN black hole, but the damping time can be longer than the maximum damping time of the RN black hole. When the EGB parameter is negative, the oscillating frequency can be higher than the maximum oscillating frequency of the RN black hole and the damping time shorter than the minimum damping time of the RN black hole.
    \item%[3)]
    When the EGB parameter is positive, the charged 4D EGB black hole cannot be distinguished from the RN black hole by means of a shadow radius. Nonetheless, the shadow radius of the charged 4D EGB black hole will be smaller than the minimum shadow radius of the RN black hole when the EGB parameter is negative.
\end{itemize}
%Through the above conclusions, we will hopefully distinguish the RN black hole of general relativity from the charged black hole of 4-dimensional EGB theory in the experiment, so as to verify the correctness of the 4-dimensional EGB theory. At the same time, we can judge the value range of EGB parameter $\alpha$ in the 4-dimensional EGB theory according to the deviations of observable measurements such as temperature, the quasi-normal mode and the black hole shadow, so as to verify whether the spacetime completeness and transformation rationality of the 4-dimensional EGB theory can be satisfied.

\section*{Acknowledgments}

The authors would like to thank X.-C. Cai, C. Lan, and Y.C. Ong for useful discussions.
This work was supported in part by the National Natural Science Foundation of China under grant Nos. 11675081 and 12175108.
The authors would like to thank the anonymous referee for the helpful comments that improve this work greatly.

%\newpage
\section*{Appendix}
\appendix

\section{Derivation of the relative difference  \label{appendix:relative_difference}}
\setcounter{equation}{0}
\renewcommand\theequation{A.\arabic{equation}}
The derivation is based on {\em Mathematical Theory of Black Holes} by Chandrasekhar~\cite{AD1}.

For a general spherically symmetric spacetime, the Lagrangian of a free neutral particle is
\begin{equation}
\mathscr{L}=\frac{1}{2}\left\{-f(r)\dot{t}^2+[f(r)]^{-1}\dot{r}^2+r^2\dot{\theta}^2+(r^2 \sin^2\theta)\dot{\varphi}^2\right\},
\end{equation}
where the dot represents the derivative with respect to the affine parameter $\tau$. The corresponding canonical momenta read
\begin{equation}
\begin{split}
p_t:=-\frac{\partial \mathscr{L}}{\partial \dot{t}}=f(r)\dot{t},\qquad p_r:=\frac{\partial \mathscr{L}}{\partial \dot{r}}=[f(r)]^{-1}\dot{r},\\
p_\theta:=\frac{\partial \mathscr{L}}{\partial \dot{\theta}}=r^2\dot{\theta},\qquad p_\varphi:=\frac{\partial \mathscr{L}}{\partial \dot{\varphi}}=r^2\dot{\varphi} \sin^2\theta.
\end{split}
\end{equation}
The resulting Hamiltonian takes the form, 
\begin{equation}
\mathscr{H}=-p_t\dot{t}+(p_r\dot{r}+p_\theta\dot{\theta}+p_\varphi\dot{\varphi})-\mathscr{L}=\mathscr{L}.
\end{equation}
The constancy of the Hamiltonian gives rise to
\begin{equation}
\mathscr{H}=\mathscr{L}=const.
\end{equation}
By rescaling the affine parameter $\tau$, we can take this constant to be $-\frac{1}{2}, 0, \frac{1}{2}$, corresponding to the time-like, null, and space-like cases, respectively. Moreover, the Hamiltonian canonical equations lead to the following integrals of motion,
\begin{equation}
\frac{dp_t}{d\tau}=\frac{\partial\mathscr{L}}{\partial t}=0,\qquad \frac{dp_\varphi}{d\tau}=-\frac{\partial\mathscr{L}}{\partial \varphi}=0.
\end{equation}
Considering the motion of a particle on the equatorial plane, $\theta=\pi/2$, we can define the energy per unit mass $E$ and angular momentum $L$ as
\begin{equation}
p_t=const.\equiv E,\qquad p_\varphi=r^2\frac{d\varphi}{d \tau}=const.\equiv L,\qquad p_\theta=r^2\dot{\theta}=0.
\end{equation}
So the Lagrangian of the particle can be written as 
\begin{equation}
2\mathscr{L}=-\frac{E^2}{f(r)}+\frac{\dot{r}^2}{f(r)}+\frac{L^2}{r^2}=-1.
\end{equation}
For free particles in the radial motion, we know $L=0$ and $E^2=1$. Thus, the square of velocity associated with  $\tau$ of a free neutral  particle is
\begin{equation}
v^2=\left(\frac{dr}{d\tau}\right)^2=1-f(r).
\end{equation}
According to the definition of the relative difference, we obtain
\begin{equation}
Diff\equiv\frac{v_\alpha^2(r)-v_{\rm RN}^2(r)}{v_{\rm RN}^2(r)}=\frac{f_{RN}(r)-f_\alpha(r)}{1-f_{RN}(r)},
\end{equation}
which is just Eq.~(\ref{relative_difference}) when $f_\alpha(r)$ (Eq.~(\ref{metric})) and  $f_{RN}(r)$ (Eq.~(\ref{metric_RN}))  are considered.

\end{document}